\documentclass[pdflatex,sn-mathphys-num]{sn-jnl}

\usepackage{graphicx}%
\usepackage{nicefrac}
\usepackage{tabularx}
\usepackage{microtype}
\usepackage{multirow}%
\usepackage{amsmath,amssymb,amsfonts}%
\usepackage{amsthm}%
\usepackage{mathrsfs}%
\usepackage[title]{appendix}%
\usepackage{xcolor}%
\usepackage{placeins} 
\usepackage{textcomp}%
\usepackage{manyfoot}%
\usepackage{booktabs}%
\usepackage{algorithm}%
\usepackage{algorithmicx}%
\usepackage{algpseudocode}%
\usepackage{float}           
\usepackage{listings}%

\theoremstyle{thmstyleone}%
\theoremstyle{thmstyletwo}%

\theoremstyle{thmstylethree}%

\begin{document}

\title[Article Title]{Poverty traps are rare, but trappedness isn't}


\author*[1]{\fnm{Isaak} \sur{Mengesha}}\email{i.a.mengesha@uva.nl}

\author[1]{\fnm{Nik} \sur{Brouw}}\email{n.w.brouw@uva.nl}
\author[1,2]{\fnm{Peter} \sur{Sloot}}\email{p.m.a.sloot@uva.nl}
\author[3]{\fnm{Egbert van} \sur{Nes}}\email{egbert.vannes@wur.nl}
\author[3]{\fnm{Marten} \sur{Scheffer}}\email{marten.scheffer@wur.nl}
\author*[1]{\fnm{Debraj} \sur{Roy}}\email{d.roy@uva.nl}

\affil*[1]{\orgdiv{Computational Science Lab, Informatics Institute}, \orgname{Faculty of Science}, \orgaddress{\street{Science Park 900}, \city{Amsterdam}, \postcode{1098 XH}, \state{North Holland}, \country{The Netherlands}}}

\affil[2]{\orgdiv{Complexity Science Hub}, \orgname{} \orgaddress{\street{Metternichgasse 8}, \city{Vienna}, \postcode{10587}, \country{Austria}}}

\affil[3]{\orgdiv{Aquatic Ecology and Water Quality Management Group}, \orgname{Wageningen University}, \orgaddress{\street{P.O. Box 47}, \city{Wageningen}, \postcode{6700 AA}, \country{The Netherlands}}}


\abstract{
The persistence of poverty is not well explained by who is poor. We argue the relevant object of measurement is trappedness—expected escape time from deprivation—which varies systematically across institutional environments and is invisible to standard poverty indices. Using Markov chains estimated on twenty years of longitudinal data from 27 European countries, we show that countries with identical deprivation rates differ in escape times by up to fourfold. These differences are not explained by household characteristics alone: exogenous shocks reshape welfare landscapes differently across countries, with divergence tracking welfare regime architecture rather than household composition. The mechanism is behavioural: health constrains a household's capacity to convert income gains into durable welfare improvement. Income transfers without health improvement fail to reduce poverty-return risk; combined interventions are super-additive across 28 countries, and the gap widens with transfer size. These findings dissolve the long-running poverty trap debate—studies that rejected traps measured the wrong dimension; studies that found them captured one projection of a multidimensional dynamic process. Trappedness is continuous, multidimensional, and institutionally shaped.
}


\maketitle
\section{Introduction}\label{sec1}

Eradicating extreme poverty remains a central challenge for the global economy. In 2023, 691 million people lived in extreme poverty—a return to pre-pandemic levels after temporary progress \cite{yang2022contribution,messerli2019global,mahler2022impact}. More troubling still, progress in poverty reduction had been slowing for over a decade before COVID-19 disrupted global economies \cite{angelov2023covid,batabyal2021covid}. This deceleration reignites fundamental questions about measurement: arbitrary cut-offs, weighting schemes, the choice between dashboards and indices, and the trade-offs these choices entail \cite{kakwani2025multidimensional}. The sensitivity of poverty statistics to measurement is not novel. Applying established measures to several countries' data reveals little consistency in who counts as poor \cite{pu2024poverty}. These critiques point to a deeper problem: an overfixation on outcomes—whether households fall above or below poverty lines—and a neglect of their ability to leave such states. This echoes longstanding calls for measures that distinguish chronic from transient poverty \cite{carter2013economics} and that handle the covariance of multiple relevant dimensions appropriately \cite{Ravallion1996IssuesIM}.\\

The problem of measurement feeds directly into the ongoing challenge of identifying poverty traps. Poverty traps are typically defined as self-reinforcing mechanisms that lock households into persistent deprivation. This definition implicitly requires two ingredients: a unified notion of deprivation or welfare, and an absolute notion of persistence over time. Different empirical traditions operationalize these ingredients in markedly different ways—often implicitly—variously emphasizing asset dynamics, income mobility, or static multidimensional deprivation \cite{kraay2014poverty,kakwani2025multidimensional}. As a result, decades of research have repeatedly asked whether poverty traps exist, yet produced strikingly contradictory findings \cite{kraay2014poverty}. Cross-country studies of asset dynamics find little evidence for the bimodal distributions predicted by classic poverty trap theory \cite{McKay2013HowSI}; others argue that traps are confined to marginalized regions and should not arise where social investments function and markets clear \cite{McKay2013HowSI}. In contrast, empirical evidence for poverty traps has been reported in rural settings \cite{ngonghala2017general} and in interaction with risk and shocks \cite{santos2017heterogeneous}. Barrett \textit{et al.} argue that identifying multidimensional traps is intrinsically difficult, and that the absence of evidence in single dimensions should not be interpreted as evidence of absence \cite{barrett2016well}. Consistent with this view, later work finds little evidence for urban poverty traps yet documents substantial persistence—or ``stickiness''—in poverty dynamics \cite{janz2023leaving}. Taken together, the literature alternates between rejecting poverty traps, localizing them to specific contexts, or replacing them with weaker notions of persistence. The debate has not converged despite decades of work. We argue this is because the question itself is ill-posed.\\

\begin{figure}[H]
    \centering
    \includegraphics[width=0.8\linewidth]{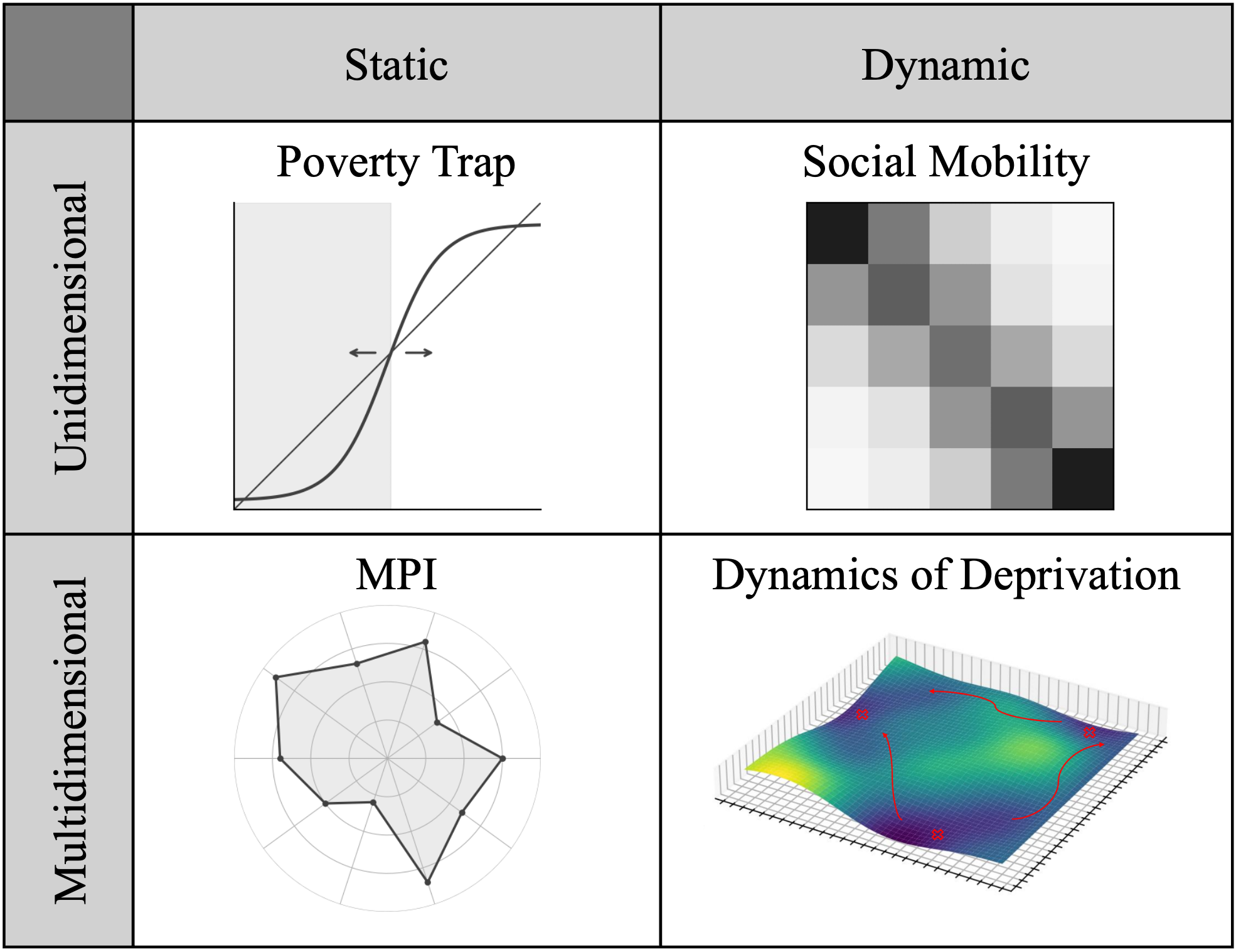}
    \caption{\textbf{From deprivation traps to trappedness.} 
    Static and unidimensional approaches treat traps as fixed objects of the environment -- i.e. the returns to investment. Static multidimensional indices capture joint deprivation -- with somewhat arbitrary thresholds -- but often see persistence as binary. Mobility matrices capture dynamics, often intergenerational income mobility, but only in a single dimension. A dynamic, multidimensional view instead represents welfare as an evolving landscape, where barriers and escape times vary continuously across income, health, and education—motivating trappedness paired with deprivation as the relevant object of measurement.}
    \label{fig:schema}
\end{figure}

The apparent disagreement reflects neither contradictory evidence nor irreconcilable mechanisms, but an ontological disagreement about what ought to be measured and how. Several well-established mechanisms—most prominently multiple financial market failures generating asset thresholds, and irreversibilities in human capital accumulation—can render poverty self-reinforcing even without cleanly identifiable equilibria \cite{barrett2016well}. Yet empirical tests of poverty traps implicitly assume these mechanisms apply uniformly and persistently across time and space, treating both poverty and traps as binary: either present or absent. The mechanisms themselves imply no such dichotomy. They imply that the difficulty of escaping deprivation varies continuously with position in a multidimensional welfare space and with the institutional environment governing risk, returns, and recovery. Smooth mobility in income can coexist with deep persistence in health and education; a household may move fluidly across income quintiles while remaining locked for decades into poor health and low human capital, or vice versa \cite{barrett2013economics}. What these mechanisms generate, then, is not a trap as an object but gradients of persistence—regions of the state space where expected escape times lengthen, recovery slows, and shocks leave durable scars. Furthermore, the underlying conditions that determine risk, volatility, and the opportunity landscape change over time. Identical states of measured deprivation at different points in time may differ sharply in how constrained future trajectories are. Whether a household is ``trapped'' is thus not a property of its current poverty status, but of the dynamic pathways available from its position in multidimensional welfare space (see Fig. \ref{fig:schema}). Unidimensional static measures cannot detect such multidimensional dynamic trappedness. The persistence of disagreement in the literature is therefore a measurement artifact, not a substantive dispute about underlying mechanisms.\\

This reframing transforms the fundamental disagreement into an empirical measurement problem: not whether traps exist, but where trappedness is high, for whom, and why. It addresses a broader limitation in poverty research—the focus on who is poor rather than why poverty persists and who is likely to remain poor under existing institutional arrangements \cite{brady2023poverty, Barrett2020FindingOB}. Abandoning the false binary of ``do poverty traps exist?'' enables progress in understanding the dynamics of deprivation. Just as acknowledging the multidimensionality of poverty opened new avenues for measurement and policy \cite{alkire2011counting}, we argue that measures of trappedness—when combined with deprivation measures such as the MPI—resolve several open problems. First, they explain why empirical studies alternately find and reject poverty traps: these results reflect different projections of a multidimensional dynamic process, not contradictory evidence. Second, they distinguish structural persistence from temporary hardship by directly measuring expected escape times rather than inferring chronic poverty from repeated snapshots. Third, they circumvent the aggregation problem of multidimensional poverty indices by replacing arbitrary weighting and cut-offs with joint dynamic analysis. Fourth, they enable policy design to target barriers and intervention sequences—not just poverty status—explaining why income-focused policies often fail to produce durable exits. These contributions rest on a specific empirical mechanism: trappedness operates through behavioural constraints---most prominently health---that mediate a household's capacity to convert resources into durable welfare gains. Institutional environments shape these constraints, making trappedness a policy variable rather than a household trait.\\

We operationalize this concept by introducing dynamic metrics that quantify expected escape time, societal rigidity, and resilience, and apply them to twenty years of longitudinal household data from 27 European countries \cite{Eurostat2023}. Longitudinal surveys now track the same households over decades, enabling dynamic analysis of poverty trajectories, yet this information remains underutilized in mainstream poverty measurement \cite{jantti2015income}. Europe serves as a limiting case: if trappedness is detectable and policy-relevant in wealthy welfare states—where classic poverty traps are widely assumed not to exist—then it captures a dimension of deprivation overlooked by existing approaches. We also explore limits on data requirements, since data quality in developing contexts is often substantially worse and high-quality longitudinal data hard to come by.\\

We uncover systematic differences in trappedness across welfare regimes despite similar levels of measured deprivation, demonstrating that trap depth reflects institutional architecture rather than immutable household characteristics. We exploit the COVID-19 pandemic as an exogenous perturbation to show that trappedness itself is not fixed. If trappedness were a stable property of household characteristics, shocks would alter who is deprived without changing the depth or structure of deprivation. Instead, we find that similar pre-pandemic welfare distributions evolved into markedly different post-pandemic landscapes depending on institutional resilience. That these divergences are associated with differences in institutional context—not household composition—indicates that trap depth reflects welfare regime architecture. Trappedness is a dynamic, context-dependent property, not a static classification attached to households.\\

These findings validate trappedness as a comparative, policy-relevant measure. Continuous measures enable analysis across countries, time periods, and welfare regimes that binary poverty classifications cannot support. By analysing multidimensional welfare dynamics, the framework identifies regions of welfare space where escape is slow and barriers are high, and highlights sequences of improvements associated with shorter expected durations in deprivation. Taken together, this reframing shifts poverty analysis from identifying who is poor at a point in time to analysing how long deprivation persists and how institutional environments shape pathways out of it.

\FloatBarrier
\section{Results}\label{sec2}
\subsection{Does 'trappedness' capture information beyond deprivation?}

\begin{figure}[h]
\centering
\includegraphics[width=1.0\linewidth]{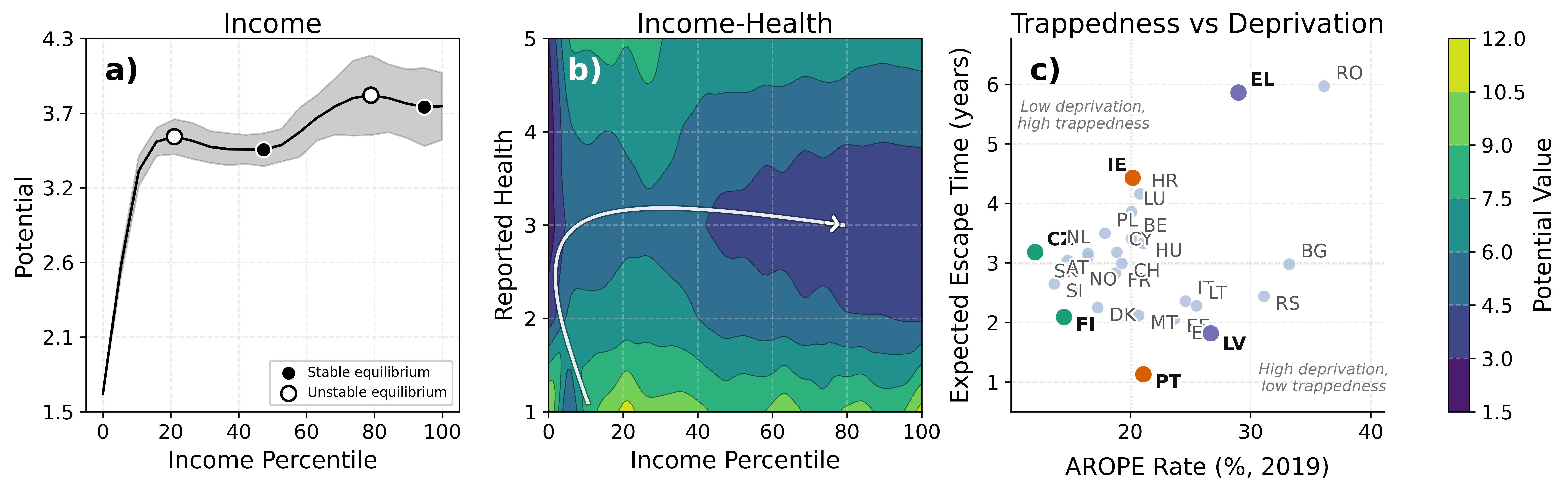}
\caption{\textbf{Multidimensional social mobility landscapes in the Netherlands.} \textbf{(a)} Potential landscape over income showing multiple equilibria: stable (filled) and unstable (open) fixed points. Shaded band: bootstrap uncertainty. \textbf{(b)} Two-dimensional landscape over income and health. Darker regions are deeper basins; contour lines denote equipotential boundaries for MFPT calculation. White arrow: stylised optimal pathway. \textbf{(c)} Expected escape time from poverty versus AROPE rate (2019, 27 EU countries). Highlighted pairs share near-identical deprivation rates yet differ in escape times by up to 4×—the two measures capture distinct properties.}
\label{fig:distinct}
\end{figure}

We estimate Markov chains over discretised welfare space from twenty years of European longitudinal household data and derive potential landscapes following the procedure described in Methods (Section \ref{sec:method}).
The one-dimensional landscape over income alone already reveals multiple equilibria: stable attractors at low, middle, and high income separated by unstable barriers (Fig. \ref{fig:distinct}a). Extending to a joint income × health topography uncovers interdependent barriers invisible in either dimension alone (Fig. \ref{fig:distinct}b). A household with moderate income but poor health sits in a deeper basin than income alone predicts.

Trappedness and deprivation are not the same quantity. Plotting expected escape time from poverty against the AROPE rate for 27 European countries in 2019 yields no tight correlation (Fig. \ref{fig:distinct}c). Country pairs with near-identical AROPE rates—Portugal and Latvia, for instance—differ in escape times by up to a factor of four. The pattern holds across the full range of European welfare regimes.
For the Netherlands specifically, the mean first-passage time from the lowest welfare state is 3.16 years pre-COVID, rising to 3.63 years during COVID (Appendix Table \ref{tab:escape_times_comparison2}). Mixing time—the horizon over which the system loses memory of initial conditions—is 42 years when estimated over income alone but drops to 16 years when income and health are analysed jointly. Adding health unlocks mobility pathways invisible in income data. In other countries the pattern reverses: adding dimensions increases mixing time, indicating that health or education constraints deepen rigidity rather than alleviate it.

\FloatBarrier
\subsection{Is 'trappedness' institutional or household-level?}

COVID-19 provides a natural experiment. A common shock hit countries with similar pre-pandemic welfare distributions. If trappedness is a household-level property, landscapes should reshuffle in composition but not topography. If it is institutional, the topography itself should change—and differently across countries.

\begin{figure}[h]
\centering
\includegraphics[width=1.0\linewidth]{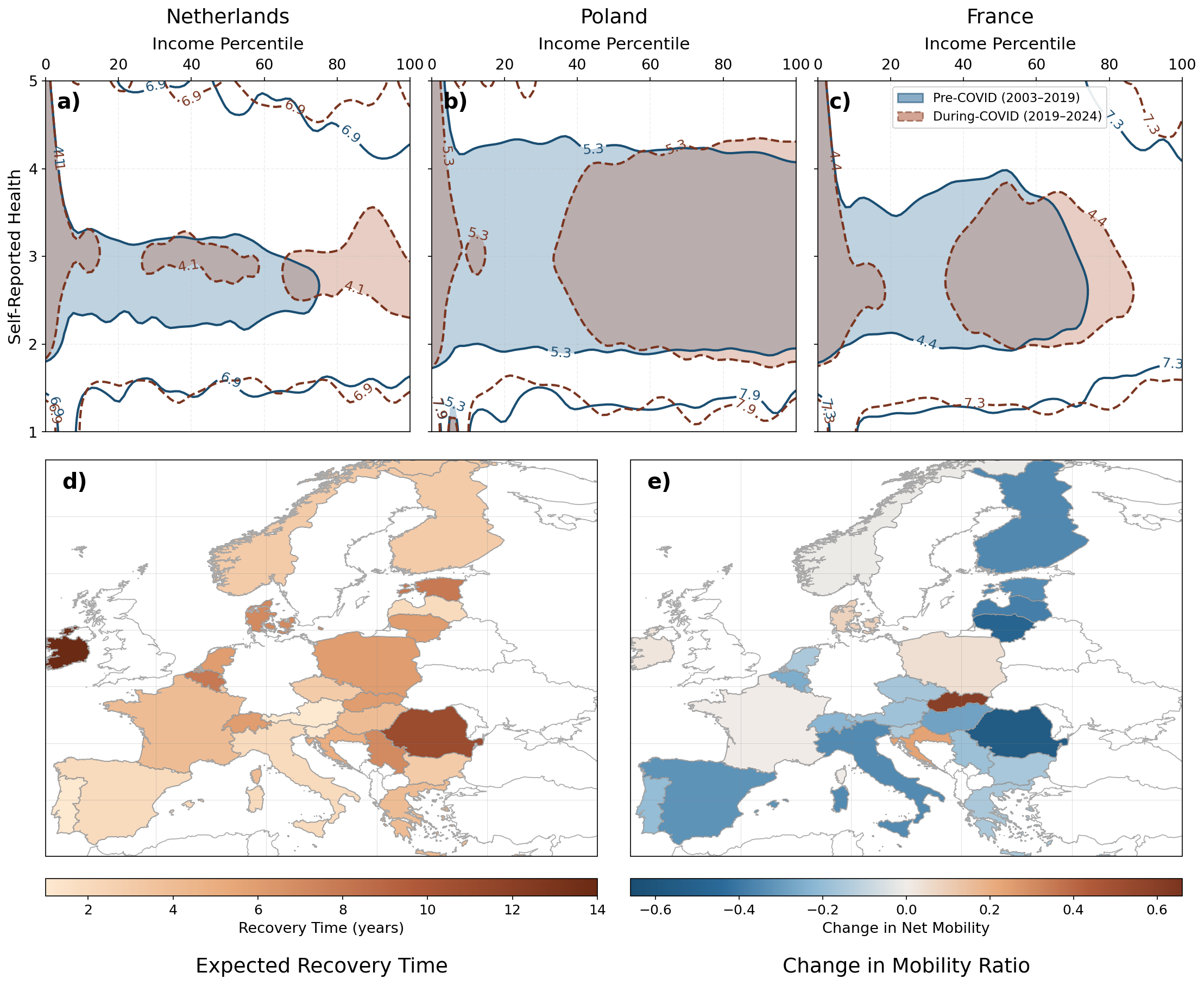}
\caption{\textbf{COVID-19 reshapes social mobility landscapes differently across countries.} \textbf{(a)--(c)} Pre-COVID (2003--2019, solid) and during-COVID (2019--2024, dashed) landscapes over income and health for the Netherlands, Poland, and France. Contour lines: equipotential boundaries for MFPT calculation; shaded regions: deepest basins. Basin contraction and displacement show that COVID differentially altered mobility pathways—lower-income groups faced steeper barriers while higher equilibria remained broadly stable. \textbf{(d)} Expected recovery times to pre-pandemic steady state, obtained by perturbing the pre-COVID steady state with the COVID-era transition matrix. Countries in white lacked data for the full period. \textbf{(e)} Normalised net mobility change during COVID; negative values indicate net downward mobility.}
\label{fig:mainpanel}
\end{figure}

The topography changed. Pre- versus during-COVID landscapes for the Netherlands, Poland, and France show that COVID reshaped the landscape itself, not just the distribution of households within it, and did so differently per country (Fig. \ref{fig:mainpanel}a–c). All three share one feature: barriers to upward mobility steepened for lower welfare states. Beyond that, the responses diverge. While the Netherlands separates into several basins of attraction, Poland and France cleanly separate into two. Across all 27 countries, lower-income, lower-health households faced disproportionately steeper barriers; upper-basin contours shifted only modestly. In several countries a new "middle" equilibrium materialised during COVID—a basin that did not exist pre-shock, trapping households pushed out of moderate welfare into a stable but precarious attractor.
The institutional gradient is clear. Countries with more generous unemployment insurance, universal healthcare, and robust education funding experienced dampened trap deepening. Countries lacking these buffers experienced the sharpest deformations.
Expected recovery times to the pre-pandemic steady state range from roughly 2 years in Scandinavia to more than 10 years in Southern and Eastern Europe (Fig. \ref{fig:mainpanel}d). Recovery time does not track the magnitude of the initial downward mobility shock (Fig. \ref{fig:mainpanel}e). Romania and the Baltic states experienced comparable net downward mobility to several Western European countries but face recovery horizons three to four times longer. MFPT ratios show downward mobility pressures intensified most in Latvia, Romania, Italy, and Spain. Similar pre-pandemic distributions diverged into markedly different post-shock topographies depending on institutional context.
One caveat: we do not formally construct a counterfactual for pandemic-absent evolution. Some landscape drift is visible in pre-COVID sub-periods (e.g. 2003–2011 vs. 2011–2019). The COVID signal exceeds this pre-existing drift in magnitude and is concentrated in the low-welfare basin, but we cannot fully rule out that part of the observed reshaping reflects secular trends accelerated rather than caused by the pandemic.

\FloatBarrier
\subsection{Behavioural limits on reducing 'trappedness'—evidence from health}

Two counterfactual analyses across 28 European countries test whether income and health improvements interact or operate independently. In each, households are moved from the lowest welfare state to the 25th percentile in income, health, or both, and five-year poverty-return risk is compared.

\begin{figure}[H]
\centering
\includegraphics[width=1.0\linewidth]{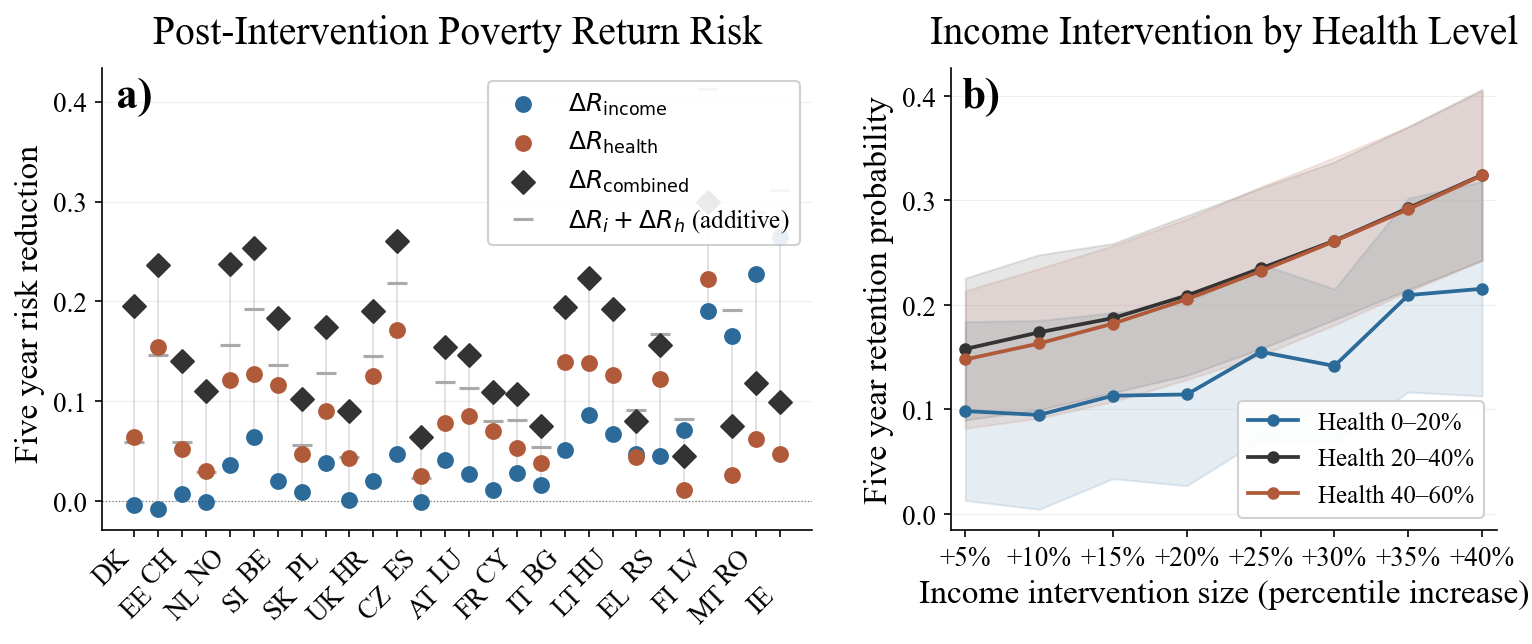}
\caption{\textbf{Health constrains the returns to income transfers.} \textbf{(a)} Five-year poverty-return risk reduction $(\Delta R
)$ for income-only, health-only, and combined interventions across 28 European countries. Households are moved from the lowest welfare state to the 25th percentile in each dimension. Countries ordered by super-additivity (grey tick): the degree to which the combined effect exceeds the sum of its parts. \textbf{(b)} Retention probability above median welfare after income boosts of varying size, stratified by starting health quintile. Better initial health yields higher retention at every transfer size, and the gap widens as transfers grow. Shaded bands: $\pm$1 s.d.\ across countries.}
\label{fig:intervention}
\end{figure}

In the majority of countries, the combined intervention exceeds the sum of income-only and health-only effects (Fig. \ref{fig:intervention}a). The degree of super-additivity varies across the 28 countries. The health-only effect is consistent, yielding a risk reduction of 0.05–0.15 across nearly all countries. The income-only effect is more variable. In some countries—Denmark and Estonia stand out—income transfers to the most deprived, absent health improvement, do not reduce five-year poverty-return risk at all; the estimated effect is near zero. Health modulates the durability of income gains (Fig. \ref{fig:intervention}b). Five-year retention probability above median welfare, stratified by starting health quintile, is monotonically ordered at every income-boost size: better initial health yields higher retention. At modest boosts of +5 percentile points, the retention gap between the healthiest and least healthy strata is approximately 5–7 percentage points. At +40 percentile points, it exceeds 10 percentage points and continues to widen. Larger income transfers show more health-dependence in durability, not less.

\section{Discussion}
\label{sec:discussion}

\subsection{Theoretical Implications: Resolving the Poverty Trap Paradox}

Some find deep poverty traps; others find smooth mobility~\cite{kraay2014poverty,carter2013economics,azariadis2005poverty,kwak2011multidimensional}. Ravallion argued the disagreement stems from measurement assumptions~\cite{Ravallion1996IssuesIM, ravallion2003debate}. Barrett and Carter argued it stems from what is measured and over what horizon~\cite{barrett2006poverty, barrett2013economics, barrett2016well, dupont2025emergent}. Both are partially right. But the question as typically operationalised demands a binary answer to a phenomenon that is neither binary nor unidimensional.

The resolution lies in conflicting dimensionality. Income alone yields smooth mobility (Fig.~\ref{fig:distinct}a). Income and health jointly yield multiple attractors and barrier structure (Fig.~\ref{fig:distinct}b). Both projections are correct; ``trap rejections'' studied the wrong axis~\cite{mckay2013strong, kraay2014poverty}.

The Introduction identified three open measurement problems. First, the aggregation problem~\cite{kakwani2025multidimensional, alkire2011counting}: the MPI compresses dimensions via weights and cut-offs, whereas the landscape extracts interaction structure from joint transitions without imposing weights. Dimension selection remains a choice; relative importance for trappedness emerges from data. Second, chronic versus transient poverty~\cite{carter2013economics, jantti2015income}: standard approaches infer chronicity from spell counts or repeated snapshots, while MFPT measures expected duration directly from transition structure. The signal is in the matrix, not the discretisation; results are robust to binning (SI Section~B). Third, the reframing from ``who is poor'' to ``who is stuck, how long, through what''~\cite{brady2023poverty, pu2024poverty}: deprivation rate answers how many, escape time answers how long, dimensional decomposition answers through which constraints. This diagnostic triple replaces a single binary. The continuous metrics that enable it---MFPT (years to escape), mixing time (institutional rigidity), resilience (post-shock recovery)---replace dichotomous classifications with a quantifiable spectrum: Denmark at 2-year recovery and low mixing time; Romania at 8+ years~\cite{stojkoski2022measures}.

``Do poverty traps exist?'' is ill-posed as typically operationalised: a binary answer demanded of a continuous, dimensional phenomenon. The landscape dissolves the question into three measurable ones. Disagreement in the literature was a measurement artefact---it disappears once deprivation and trappedness are measured as the distinct objects they are.

Europe is where traps are assumed absent. That trappedness is nonetheless detectable, varies across welfare regimes, and responds to shocks confirms the framework captures a dimension existing measures miss---even in the most buffered setting. The extrapolation to low-income contexts is not automatic, however. Europe's institutional heterogeneity generates the cross-country variation we exploit; in settings with uniformly weak institutions, the framework may show uniformly high trappedness with less diagnostic power. Conversely, where deprivation is deeper, landscape gradients may be steeper and more policy-informative. Validation in low-income panels remains a priority and is the content of ongoing work (Section~\ref{sec:limitation}). If trappedness is continuous, dimensional, and institutional, interventions must target landscape topology, not poverty status. We turn to the structural basis for multidimensional intervention design.

\subsection{Structural foundations of multidimensional intervention design}

That combined interventions outperform single-dimension transfers is settled empirically (Banerjee et al.\ 2015). The mechanism is not. Cash transfers have long been prioritised on the basis of liquidity constraints and immediate needs, but cannot overcome structural barriers alone. Our explanation: intermediate basins of attraction exist between the bottom and median welfare. Single-dimension interventions deposit households \emph{into} these basins, not through them.

The mechanism runs as follows. Income alone does not reshape the basin a household occupies. Health shocks pull the household back. Addressing health alongside makes income gains anchoring rather than buffering~\cite{baird2014conditional}---a dynamic that connects to documented health--poverty trap loops~\cite{fu2024inspecting}. This process is invisible to unidimensional evaluation: income improves, but multidimensional retention is weaker than expected. Fig.~\ref{fig:intervention}a confirms the pattern: income-only and health-only interventions each produce modest risk reductions; the combined effect exceeds their sum in most countries. Fig.~\ref{fig:intervention}b shows that health modulates income retention at every intervention size, with the gap widening as transfers grow---health binds more, not less, as income rises. This is the dose--response curve the graduation literature lacks.

Poor health constrains labour market participation, increases exposure to income shocks through illness and caregiving, and erodes human capital accumulation. Income gains without health improvement produce a temporary lift; the trajectory bends back toward the low-welfare basin. Health, in this framing, governs a household's absorptive capacity---its ability to convert external support into durable welfare improvement. Super-additivity across 28 countries reflects this: dimensions interact because behavioural pathways out of poverty require capacities spanning both.

Super-additivity varies across the 28 countries, meaning the optimal bundle is context-dependent, not universal. Not every context needs the full graduation package: in countries where effects are additive, simpler transfers suffice. Familiar failures become predictable from basin structure: school feeding fails without health~\cite{banerjee2007remedying}; health infrastructure fails without income. The paradigm inverts: diagnose the topology first, then design---rather than bundle everything and evaluate post hoc. Dimensions interact through shared topology, not additive channels. The contribution is not the finding that combinations work, but its structural basis---predicting which combinations, how much, and for whom from observational data alone. This complements RCTs rather than replacing them.

\subsection{Institutional Vulnerability and Shock-Responsive Traps}

Trap architecture is not fixed. COVID as a natural experiment shows that identical pre-shock distributions diverged post-shock depending on institutional context (Fig.~\ref{fig:mainpanel}d,e).

What the transition matrix captures is neither purely institutional nor purely household-level---it reflects the \emph{effective dynamics} of households navigating a given institutional environment. The landscape is the joint product of both. But the COVID test is informative: household composition changed little over 2019--2024, while institutional responses varied dramatically. The divergence in post-shock landscapes is therefore more plausibly attributed to institutional variation than to shifts in household characteristics. This does not rule out household-level heterogeneity. Unmeasured attributes---age, networks, risk aversion, human capital---shape how individual households navigate the landscape. What we measure is the population-averaged effective potential: the landscape as experienced by the typical occupant of each state.

The causality claim is circumscribed. Trap depth covaries with welfare regime architecture and responds to institutional perturbation. COVID altered labour markets, mortality patterns, and social networks alongside policy responses, so we cannot cleanly isolate institutional channels. The key test, however, is not causal identification of a single channel but rejection of the household-fixed alternative: if trappedness were an immutable household property, landscape topography would not reshape differentially across countries under a common shock. It does, ruling out the strong household-fixed hypothesis without establishing clean institutional causation.

A bifurcation-theory framing sharpens the point~\cite{batabyal2021covid,scheffer2001catastrophic}: near critical points, small differences in institutional buffering produce qualitatively different long-term outcomes. The institutional environment sets the topology; households navigate it. Fragile regimes saw new attractors emerge; robust regimes saw dampened trap deepening. Recovery times vary from 2 to 14 years (Fig.~\ref{fig:mainpanel}d) and correlate with institutional variables---social protection spending, healthcare access---not pre-shock deprivation levels. The specific buffers that matter are unemployment insurance, public healthcare, and education funding. Countries with these entered COVID with welfare distributions less vulnerable to bifurcation.

This reframes resilience theory~\cite{johnstone2023shocks} into three layers: ex-post response, ex-ante preparation, and---decisively---the prior institutional architecture that determines whether bifurcation-prone critical points exist at all. Our evidence points to this third layer as decisive. Combined with the intervention results in Section~2.3, the implication is that the effective landscape is not static. If institutional context shapes the topology, then shocks that alter institutions also alter which interventions work. Pre-COVID optimal bundles may fail in post-shock landscapes with different basin structures. Climate adaptation, green transitions, and pandemic preparedness are anticipated perturbations that will reshape welfare landscapes; policy must account for this malleability. Welfare state investments flatten barriers and widen escape pathways---they shape the topology that households navigate. This is infrastructure, not consumption~\cite{fu2024inspecting}.

\subsection{Limitations and the Measurement Frontier}\label{sec:limitation}

Our approach faces constraints that warrant candid discussion. First, self-reported health is endogenous to income, employment, and psychological state. What we interpret as "health constraining income retention" may partly reflect reverse causation or common-cause confounding. The super-additivity result (Fig. 4a) is robust to this concern—it holds whether health causally constrains income or merely proxies for an unmeasured capacity variable—but the mechanistic interpretation (health as absorptive capacity) requires stronger identification than our observational design provides. Objective health measures in future panels would sharpen this distinction. Second, longitudinal household surveys, while unprecedented in coverage and length, remain limited in temporal resolution.
EU-SILC interviews households annually; monthly or real-time data might reveal additional dynamics obscured by aggregation. Third, our discrete state binning (necessitated by feasible transition matrix estimation) sacrifices distributional detail. Threshold precision matters for policy targeting. Fourth, the Markov assumption is memoryless by construction, while classic poverty trap theory emphasises path-dependence and cumulative scarring. Our framework measures where persistence is high—expected residence times, barrier heights—but cannot distinguish genuinely self-reinforcing feedback loops from slow diffusion through a high-friction region of state space. We measure trappedness as an empirical pattern, not as a confirmed mechanism. Establishing which basins reflect true feedback traps versus slow-mixing dynamics requires complementary causal designs (e.g. RCTs). Finally, and most fundamentally, our framework is data-hungry, requiring decades of longitudinal coverage and large sample sizes. Most developing country contexts lack such resources. Methodological adaptations—Bayesian approaches tolerating sparse data, transfer learning from similar contexts, synthetic panel construction—may extend applicability, but these remain frontier methods requiring validation.









\section{Methodology and Data}\label{sec:method}
\begin{figure}[h!]
\centering
\includegraphics[width=0.8\linewidth]{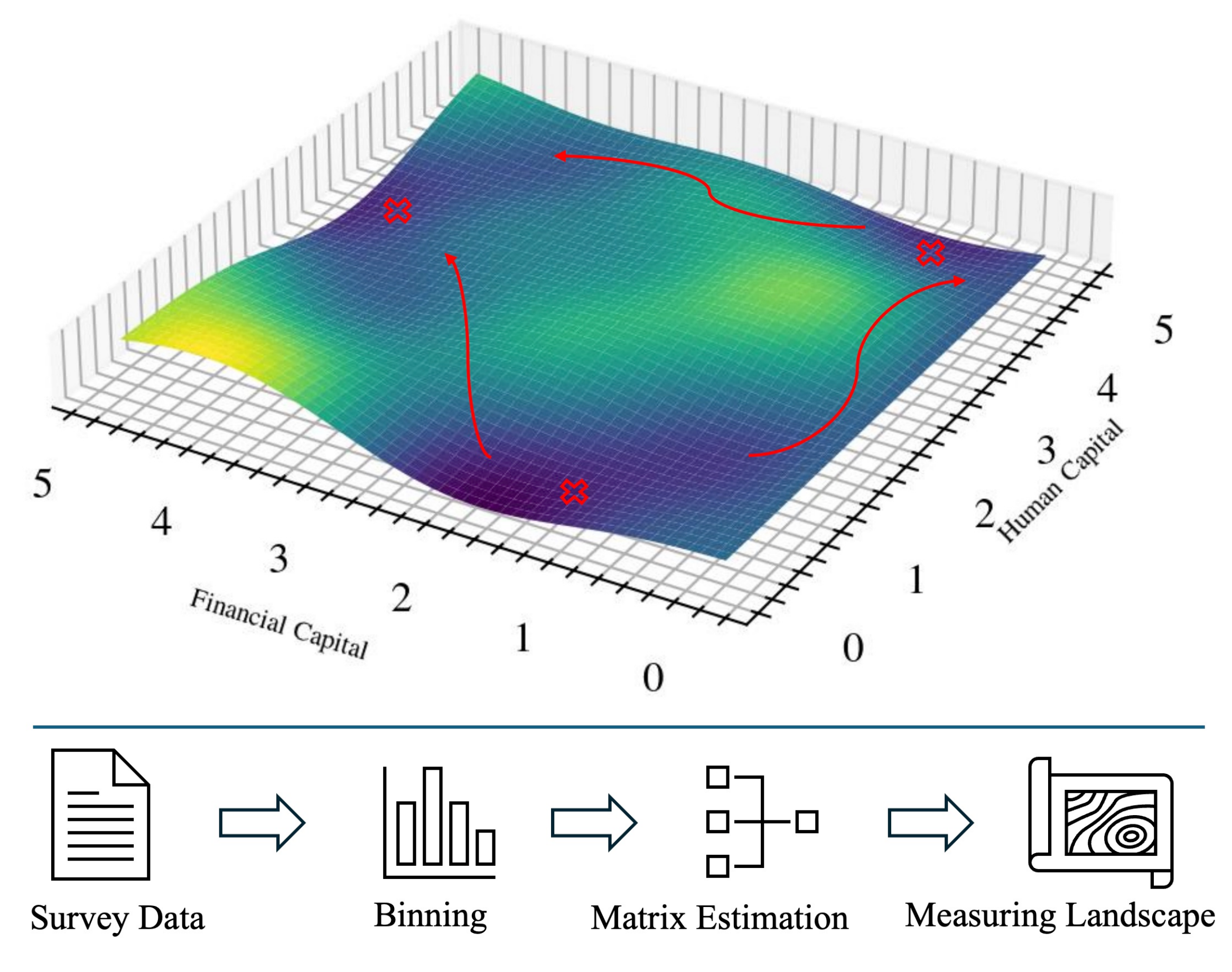}
\caption{\textbf{A 2-dimensional economic landscape} We define an economic landscape as a multidimensional, dynamic representation of the distribution and transitions of welfare states within a population over time, integrating key dimensions such as income, education, and health. It is conceptualized as a topographical map of social and economic opportunity, where stable welfare equilibria (red cross) constitute “basins,” and barriers or poverty traps form ridges and valleys. Transitions between these states, modelled as stochastic processes via Markov chains, capture the temporal and synergistic interactions shaping upward and downward mobility. Navigating between different fixed points (red crosses) raises the question of optimal paths. Alleviation pathways (red arrows) out of poverty focusing on only one dimension might neglect more cost-effective pathways. The methodological approach: longitudinal survey micro-data is binned into discrete states, transition matrices are estimated and subsequently analysed for resulting dynamics.}
\label{fig:example}
\end{figure}

In this section, we outline the methodology used to analyse social mobility through a Markov chain framework. Our approach follows a structured pipeline that systematically transforms empirical observations into an ``economic landscape" (see Fig. \ref{fig:example}). As a first step, we select measurement dimensions that capture key aspects of social mobility. We focus on self-reported health, income, and education as initial dimensions, given their well-documented relevance, although the approach remains agnostic to these choices \cite{tilak2002education}. These dimensions are discretized into a finite state space, where the choice of bin size is constrained by dataset size—smaller datasets require larger bins to ensure statistically significant transition estimates with low variance (see Supplementary Information (SI) section B in  for estimations).

\subsection{Estimating a transition matrix and potential energy landscape}
\label{subsec:landscape}
As we collect data on households' movement in some finite state space $\mathcal{E} = \{1, 2, \dots, N \}$, we effectively collect sequential data on discrete transitions in said space.
Thus each household generates a sequence of transitions $X_{i,t} \sim P_{i,j}, \quad i,j \in \mathcal{E}, \quad t \in \mathbb{T}$.
Our goal is to find a data-driven generator for the observed stochastic variable $X_{i,t}$ with an interpretable mechanism. We propose to approximate that mechanism by a Markov chain, as is common in various modelling approaches \cite{chakraborti2011econophysics,mccall1971markovian,bernstein2018poverty}.
The problem is to estimate the underlying transition matrix $P$ of the Markov chain based on the observed transitions. Our estimation of $P$ enables measurements about the stationary distribution $\pi$, given no further perturbations, as well as mixing times that closely relate to social mobility \cite{stojkoski2022measures}. The limits of these approximations have been explored by others \cite{mcfarland1970intragenerational}. In this work, we resorted to the MLE approach by Schreiber \textit{et al.}\cite{Schreiber2017}. However, in principle, this work does not hinge on the concrete matrix estimation method, as alternatives would yield similarly practical results. Subsequently, we compute the resulting potential $\Phi_{*}$ from the corresponding steady state $\pi_{*}$ of the estimated transition matrix $P_{*}$. Inspired by statistical mechanics, we compute the potential using a ``Boltzmann-like" informational value of observing this state, 
\begin{equation}
\Phi_{*}(i) = -\log \pi_i^*.
\end{equation}
However, this requires that the matrix be irreducible, which is not guaranteed due to stochastic noise from observed data. To address this, we add comparatively small noise terms to each entry of $P$, ensuring its irreducibility. Furthermore, the method presumes that the estimated transition matrix $P_{*}$ can be expressed as the gradient of an underlying scalar potential, i.e., its curl vanishes. We test this by computing the curl and consistently find its magnitude to be below $1\%$ (see Appendix table A.4).

\subsection{Measures of Markov Chains}
\label{subsec:measures}
We analyse the transition matrix of the underlying Markov chain to capture both long-run behaviour and expected duration in specific states. A key concept is the mixing time \(\tau\), which indicates how quickly the Markov chain converges to its steady-state distribution regardless of the initial state. Formally,
\[
\tau(\epsilon) 
= \min \{ t \,\mid\, \| P^t(x, \cdot) - \pi(\cdot) \|_p < \epsilon,\ \forall x \},
\]
where \(\pi(\cdot)\) is the stationary distribution, \(P^t\) is the \(t\)-step transition probability, and \(\epsilon\) is a small tolerance. Since \(\tau\) can be unbounded and inversely related to mobility, it does not directly meet the criteria proposed by Stojkoski \textit{et al.} \cite{stojkoski2022measures}. To address these requirements, we often apply a bounded, positively oriented transformation such as
\[
\tau_{\text{mix}}= 1 - e^{-k\,\tau},
\]
which remains in the interval \((0,1)\). In this form, lower mixing times (indicating faster convergence and higher mobility) yield larger values, aligning with the principle that a good mobility measure should be normalized, monotonic, period-dependent, and distribution-dependent. Mixing times in Markov chains can be applied to identify economic traps and assess economic mobility. In the context of economic mobility, mixing times can help analyse how quickly individuals or groups transition between different economic states, such as income levels or employment statuses. This analysis can reveal the persistence of economic inequality and the likelihood of moving out of poverty. Additionally, understanding mixing times can identify economic traps, where certain states, like poverty, are absorbing or difficult to escape, highlighting areas where policy interventions may be needed to enhance mobility and reduce inequality.

Further, this paper explores mobility, raising a fundamental question about the expected time required to transition out of a given state, with a specific focus on the lowest income level. Therefore, we consider the mean first passage time (MFPT), which provides the expected time to enter a particular state for the first time. Specifically, if \(T_{ij}\) is the random variable denoting the first passage time from state \(i\) to state \(j\), then
\[
M_{ij} = \mathbb{E}[T_{ij}].
\]
In the context of persistent poverty, a high MFPT from a poverty state to a non-poverty state indicates that individuals remain trapped in poverty for extended periods, while a shorter MFPT suggests quicker exits. This distinction helps identify structural barriers to upward mobility across regions or over time. We can generalize MFPT to measure the time it takes to move from a set of states \(A\) to another set \(B\). If \(T_{A \to B}\) is the first passage time to any state in \(B\) starting from any state in \(A\), then
\[
M_{A \to B} = \mathbb{E}[T_{A \to B}].
\]
This is particularly useful when “poverty” is itself a collection of states, allowing us to capture the overall duration before individuals exit a broader category of economic hardship.

\subsubsection{Shorrocks Index}
\label{shorrocks}
We can measure mobility across countries and over time using the Shorrocks index, which summarizes persistence in transition matrices where each entry \( p_{ij} \) represents the probability of moving from state \( i \) to state \( j \). The diagonal elements \( p_{ii} \) indicate state retention, with higher values signalling lower mobility. The Shorrocks index condenses this information into a single measure:

\[
M_S = \frac{N - \sum_{i=1}^{N} p_{ii}}{N-1}
\]

where \( M_S = 0 \) if all individuals remain in their initial state (no mobility) and \( M_S = 1 \) if no one does (complete mobility). As a normalized measure, it enables direct comparisons across countries and time periods. A declining \( M_S \) suggests increasing rigidity, while a rising index indicates greater fluidity. By computing \( M_S \) across different contexts, we assess how mobility patterns evolve, identifying structural persistence and responses to economic shocks or policy reforms. Lower values may indicate entrenched inequalities, informing targeted interventions. The Shorrocks index provides a simple and interpretable summary of persistence but does not capture movement direction or magnitude beyond whether an individual transitions out of their initial state. As such, it serves as a useful first-pass measure, complementing more detailed mobility analyses.

\subsubsection{Estimating Recovery Time}
\label{recovery}
To assess the resilience of social mobility to exogenous shocks, we estimate the recovery time following perturbations. A pre-COVID transition matrix $P^{\text{pre}}$ is inferred, and a COVID-period matrix $P^{\text{covid}}$ is estimated to capture pandemic-induced shifts. We simulate the shock by applying $P^{\text{covid}}$ three times to the steady-state distribution of $P^{\text{pre}}$, yielding
\begin{equation}
\pi^{*} = \pi (P^{\text{covid}})^3.
\end{equation}
Recovery is evaluated by iteratively applying $P^{\text{pre}}$ to $\pi^{*}$ and tracking its divergence from the original steady state using KL divergence.
The recovery time $t^*$ is the first iteration where $D_{\text{KL}}(\pi\,\|\,\pi^{(t)}) < \epsilon $ falls below a predefined threshold $\epsilon$, indicating system stabilization.

\subsection{Mechanistic Validation}
We validate the Markov Chain framework through rigorously evaluating the validity of underlying assumptions. We verify that the state transitions are independently and identically distributed (i.i.d.) (see Appendix section A.1) and then conduct tests for Markovian assumption (see Appendix section A.4), time homogeneity, and stationarity (see Appendix section A.3). Additionally, we compute the curl to assess detailed balance, providing insight into the system’s reversibility and equilibrium properties (see Appendix Table A.4). In the following sections, we analyse the the resulting Markov chains and focus on three primary dimensions: total income, self-reported health, and highest education level, inspired by established multidimensional poverty indices \cite{alkire2013multidimensional,alkire2015multidimensional}. Understanding social mobility and its interplay with external shocks is crucial for fostering human development. Among potential natural experiments, we examine the impact of COVID-19 pandemic on different countries in Europe, which has been widely identified as a major shock to social mobility \cite{balestra2022current}.

\subsection{Data \& Sources}
\label{sec:data_sources}

The primary dataset employed in this analysis is the EU Statistics on Income and Living Conditions (EU-SILC) provided by EUROSTAT \cite{Eurostat_EU_SILC_2024}, spanning over twenty years of household survey data across numerous European countries. Through its annual longitudinal design, EU-SILC enables tracking of the same households (identified via a unique \textit{household-id}) and offers repeated measurements of key indicators such as income, education, and self-reported health. Income was subject to corrections via a world bank dataset on longitudinal PPP values \cite{WorldBank_PPP_2024}. These variables are first binned (either equidistantly or by percentiles) to form the discrete state space for our Markov chain modelling. Any transition involving missing values in one or more dimensions is excluded, ensuring that only valid and complete data points contribute to our transition estimates. Countries with limited longitudinal coverage or high levels of missing data, such as Germany prior to 2019, are also omitted to maintain model reliability. For more information consult section E in the supplementary information. 

Despite the broad coverage of EU-SILC, potential biases may arise from self-reported variables and under-reporting of income. To help validate our assumptions, we cross-reference these data with national-level statistics from the Central Bureau of Statistics (CBS), where available \cite{DHSData}. The findings of the two data sources generally align, indicating that the transition probabilities derived from EU-SILC offer a robust depiction of social mobility and welfare dynamics within the sampled European populations.


\section*{Acknowledgements}
This work was supported by the University of Amsterdam under grant WBS C.2324.0584 (Co-producing a Social Model of Well-being with SamenWelzijn). The EU-SILC microdata were made available by Eurostat.

\clearpage
\bibliography{sn-bibliography}

\newpage
\setcounter{page}{1}

\begin{appendices}

\section{Assessing mechanistic assumptions}
\label{sec:mechanistic}
\subsection{Adjustments to panel data}
\label{iid}

Observed transitions are based on a representative sample of agents of the population under consideration. In the discrete time step $\Delta t$, each agent moves between states in $\mathcal{E}$. 
Translating this into the Markov chain framework, we have a chain \(X = \{X_{i,t} : i \in \mathcal{E}, t \in \mathbb{N}\}\). 
The chain is initialized multiple times, with each initial state \(X_{i,0}\) sampled independently and identically distributed (i.i.d.) from a distribution \(\mu\) on \(\mathcal{E}\). The distribution $\mu$ corresponds to the occupation of states by the population at large. For each initialization \(i\), we observe the state of the Markov chain at a fixed time \(\tau\), which is a positive integer. This observation represents a transition from the initial state \(X_{i,0}\) to the state \(X_{i,\tau}\).

We aim to prove that these transitions \(X_{i,0} \rightarrow X_{i,\tau}\) are independent and identically distributed (i.i.d.).

\begin{proof} 
    \textbf{Independence:} Since each \(X_{i,0}\) is sampled i.i.d. from \(\mu\), the process for each initialization \(i\) is independent of the others. By the Markov property, the probability of transitioning from the current state to any future state depends only on the current state and not on the path taken to reach that state. Formally, for any \(s, s' \in \mathcal{E}\),
    \begin{equation}
        P(X_{i,t+1} = s' | X_{i,t} = s) = P(X_{i,t+1} = s' | X_{i,1} = x_1, \ldots, X_{i,t} = s).
    \end{equation}
    Therefore, the transition \(X_{i,0} \rightarrow X_{i,\tau}\) for each initialization \(i\) is independent of the transitions in other initializations.
    
    \textbf{Identical Distribution:} The distribution of \(X_{i,\tau}\) depends on the initial state \(X_{i,0}\) and the transition probabilities of the Markov chain. Since \(X_{i,0}\) is chosen i.i.d. from \(\mu\), and assuming the transition probabilities of the Markov chain are stationary (i.e., do not change over time or across different initializations), the distribution of \(X_{i,\tau}\) is the same for each initialization. Formally, for any \(s, s' \in \mathcal{E}\),
    \begin{equation}
        P(X_{i,\tau} = s' | X_{i,0} = s) = P(X_{j,\tau} = s' | X_{j,0} = s),
    \end{equation}
    for any initializations \(i\) and \(j\). Therefore, the transitions \(X_0^{(i)} \rightarrow X_\tau^{(i)}\) in the given Markov chain are independent and identically distributed.
\end{proof}

\subsection{State-Space Discretization}
\label{space}
Before assessing the social mobility of a given population from available mobility data, it is necessary to define the state-space of interest. A common choice here is occupation, income, and education level. But in principle any recorded socio-economic measure (e.g. health) could be of interest and the approach should be agnostic to that. Furthermore, the interaction effects between different dimension should not be ignored. We first demonstrate the approach and state-space considerations in one dimension (income). We elaborate on how to expand this approach to multiple dimensions in the Supplementary Information (SI). 

Given a continuous household income distribution, denoted by the random variable $Y$, we discretize the income space into a finite set of intervals (bins), denoted by $B_1, B_2, \dots, B_K$, where $K$ is the total number of bins. Two common approaches to binning can be defined as follows:

\begin{enumerate}
    \item \textbf{Equidistant Binning}:  In this approach, the income range $[Y_{\min}, Y_{\max}]$ is divided into $K$ equal-sized intervals:
    \begin{equation}  
    B_i = \left[Y_{\min} + (i-1) \cdot \frac{Y_{\max} - Y_{\min}}{K}, Y_{\min} + i \cdot \frac{Y_{\max} - Y_{\min}}{K} \right), \quad i = 1, 2, \dots, K .
    \end{equation}
    This method focuses on capturing \textbf{absolute mobility} by tracking the changes in income levels within fixed brackets of the income distribution.

    \item \textbf{Density-based (Percentile) Binning}:  Alternatively, we can partition the income distribution based on quantiles, ensuring equal numbers of households in each bin. Let $F_Y(y)$ denote the cumulative distribution function (CDF) of $Y$. The bins are defined such that:
    \begin{equation} 
    B_i = \left[F_Y^{-1}\left(\frac{i-1}{K}\right), F_Y^{-1}\left(\frac{i}{K}\right)\right), \quad i = 1, 2, \dots, K,
    \end{equation}    
    where $F_Y^{-1}$ is the inverse CDF (also known as the quantile function). This approach is useful for studying exchange mobility, as it focuses on changes in the relative ranking of households within the distribution.
\end{enumerate}

The choice between equidistant and density-based binning depends on whether the analysis aims to capture shifts in the absolute income distribution (structural mobility) or changes in the relative income ranks (exchange mobility).

Additionally, we need to consider the trade-offs involved with the choice of resolution. The larger the bin size, the less informative the resulting transition matrix, but the higher the accuracy. Conversely, the smaller the bin size, the higher the resolution and structural information but larger bias and noise contributions. 

\begin{figure}[H]
  \centering
  \begin{minipage}{0.45\linewidth} 
    \centering
    \includegraphics[width=\linewidth]{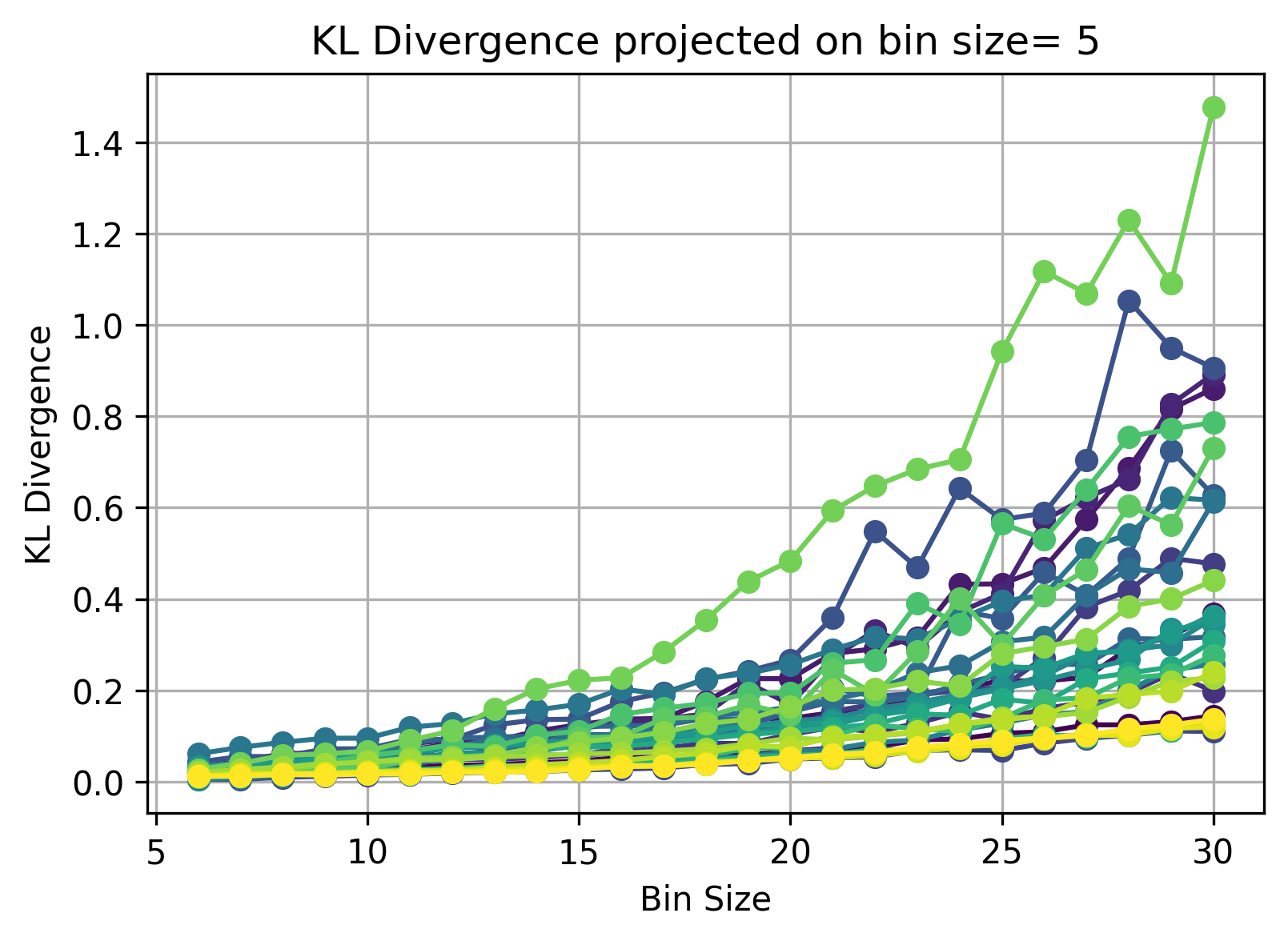} 
  \end{minipage}
  \hfill
  \begin{minipage}{0.45\linewidth}
    \centering
    \includegraphics[width=\linewidth]{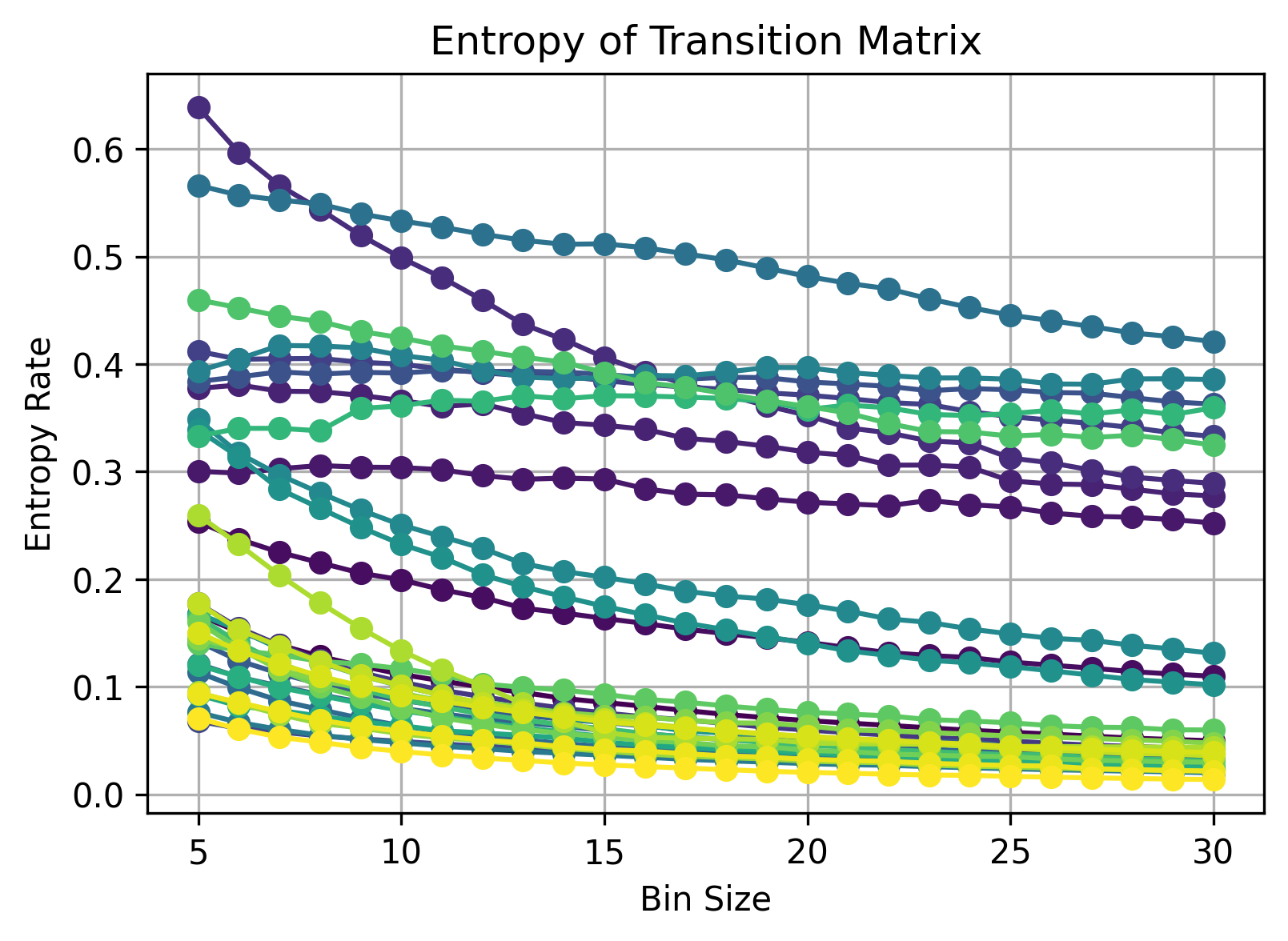}
  \end{minipage}
  \caption{We first compute the transition matrix for varying numbers of bins and density-based binning for all countries (represented by different colors). (left) The KL divergence is computed by projecting higher resolution matrices to rank=5 and comparing the distribution. (right) The entropy rate normalized by log(n), with decreasing values. This means the system reveals increasing structure with increasing resolution. }
  \label{fig:binningeffects}
\end{figure}

With the now discretized state space we can continue to estimate the underlying transition matrix. The trivial or empirical estimator $Q$ would be to simply normalize the recorded transitions. There exist various methods to obtain the underlying transition matrix of a Markov process from observed data (e.g. maximum likelihood or Bayesian approaches) \cite{sherlaw1995estimating, eichelsbacher2002bayesian}. The subsequent analysis, however, is independent of the estimation method; here we opt for a simple MLE approach.

From here we aim to  approximate the underlying stochastic process using a first-order Markov chain rests on the assumption that future states depend mostly on the present state. Although many real-world processes exhibit higher-order dependencies or long-range correlations, a first-order Markov chain often provides a useful approximation by balancing simplicity and interpretability with capturing essential dynamics. Even if a process is not strictly Markovian, the first-order approximation can be effective if the influence of past states diminishes quickly, or if transitions are primarily driven by local dependencies. The reduced model complexity compared to higher-order or non-Markovian models allows for tractable analysis and avoids overfitting, especially in systems where the goal is to capture coarse-grained behaviour. Moreover, empirical evidence often supports this approximation, particularly when the observed transition probabilities show that the immediate past state provides the most predictive power for the next state. Furthermore, many of the mobility measures deployed are related to the convergence behaviour of the Markov chain and therefore sensitive to additional assumptions of homogeneity and stationarity. Therefore we must acknowledge violations to these assumptions and estimate their impact on our results.

\subsection{Homogeneity and Stationarity}
\label{time}
Our model relies on a key simplification: individuals or households are assumed to be indistinguishable if they occupy the same state in the state space $\mathcal{E}$. This implies that their current state serves as a complete description of their relevant socioeconomic characteristics. While this simplification is common in mobility studies (see \cite{Hout2015ASO}), it inevitably overlooks certain confounders such as regional differences, which may introduce variations within states. Acknowledging this limitation, our focus remains on modeling the broader structure of mobility, understanding that higher granularity can be introduced by expanding the dimensionality of $\mathcal{E}$ to include factors such as health, education, or assets \cite{bellman1966dynamic}. However, adding such dimensions would increase the complexity of the model and may encounter issues related to the curse of dimensionality. We also choose to avoid using hidden Markov models, which are more established in financial applications \cite{bhar2004hidden}, because the design choices for such models require qualitative input from domain experts, and they introduce the risk of overfitting, particularly in settings with limited data.

\begin{figure}[H]
    \centering
    \includegraphics[width=\textwidth]{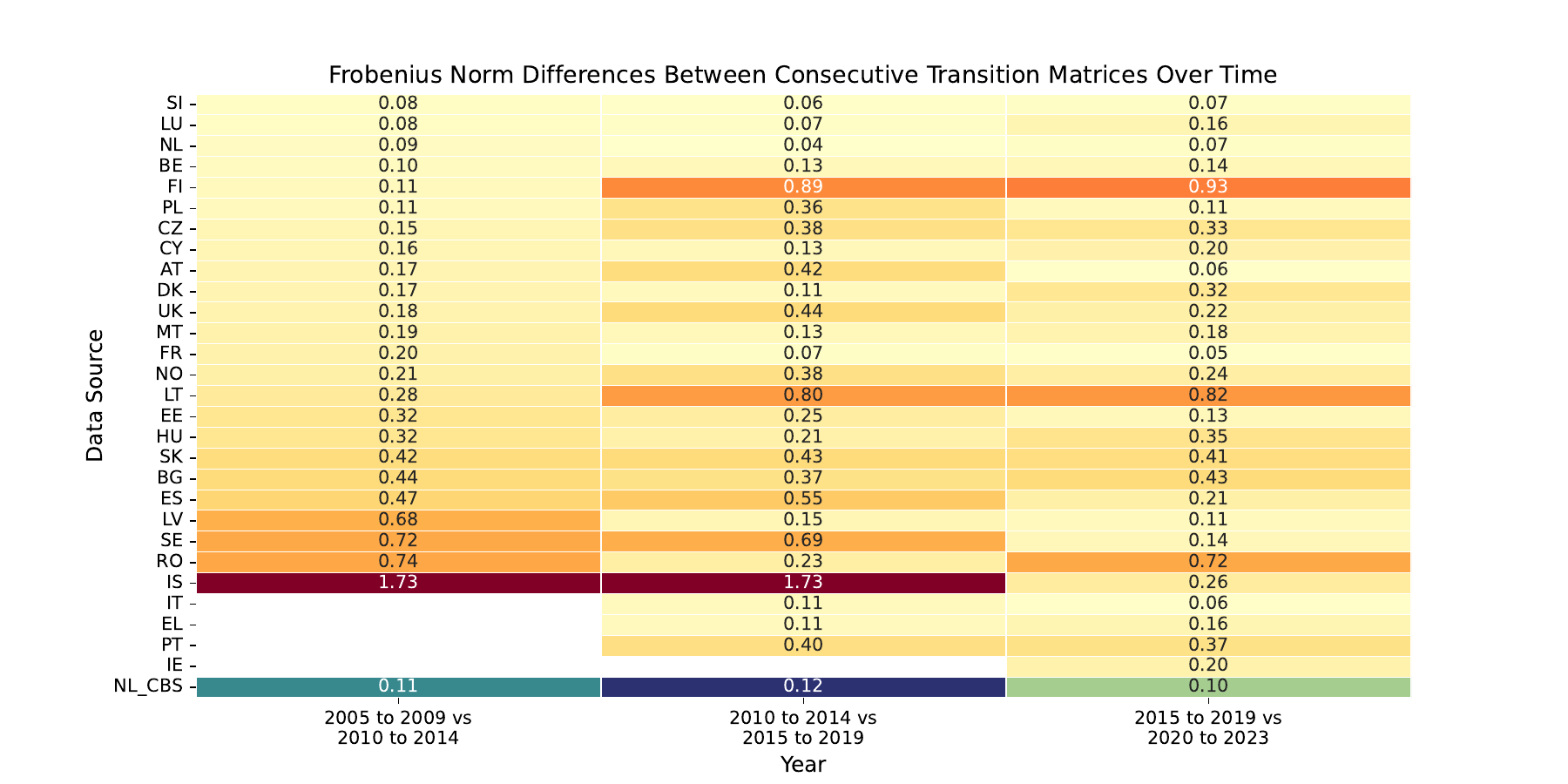}
    \caption{Computing Frobenius-norm for the method-obtained estimator for 5-year intervals. }
    \label{fig:frobenius}
\end{figure}

In the interest of simplicity, we assume that the transition rates between states are stationary, meaning they remain constant over time. This assumption, while theoretically convenient, may not hold in dynamic socioeconomic systems. To address this, we quantify the degree to which stationarity is violated in our data using the Frobenius norm of the differences in transition matrices across time intervals. In our case, the norm is approximately $0.10$, which provides a measure of how much transition probabilities deviate over time. This value will be used as an error estimate when analysing properties such as mixing time. Furthermore, while ergodicity—often crucial in stationary Markov models—is not a primary concern in our model, it’s important to note that large transient subspaces may exist, especially in growing economies. These subspaces may exhibit divergent behaviour over time, though transient states eventually become irrelevant in terms of long-term mobility outcomes. Notably, in density-based binning approaches, the presence of large transient states is unlikely.

\subsection{Memory length}
\label{memory}

The memory length of a Markov chain plays a critical role in capturing the complexity of state transitions within the modelled system. A first-order Markov chain assumes that the future state depends solely on the current state, while higher-order Markov chains incorporate dependencies on one or more previous states. These higher-order models can potentially provide a more accurate representation of systems where historical states influence current transitions. This is technically approached by augmenting the state space: for a second-order Markov chain, the state at time \(t\) is represented by the tuple \((X_{t-1}, X_t)\), and transitions occur over these composite states. The resulting process satisfies the Markov property in the augmented space. Moreover, since each trajectory is an independent sample from the same underlying Markov process, and the augmentation is a deterministic transformation applied identically to each, the set of augmented sequences preserves the i.i.d.\ structure across samples.

However, higher memory lengths come with increased model complexity, potentially leading to over-fitting, especially in data-limited scenarios. Moreover, as demonstrated in various studies, the marginal gain in predictive accuracy from extending the memory length may diminish. Thus, identifying an optimal balance between model complexity and predictive power is crucial for meaningful analysis.

\begin{figure}[H]
    \centering
    \includegraphics[width=0.8\textwidth]{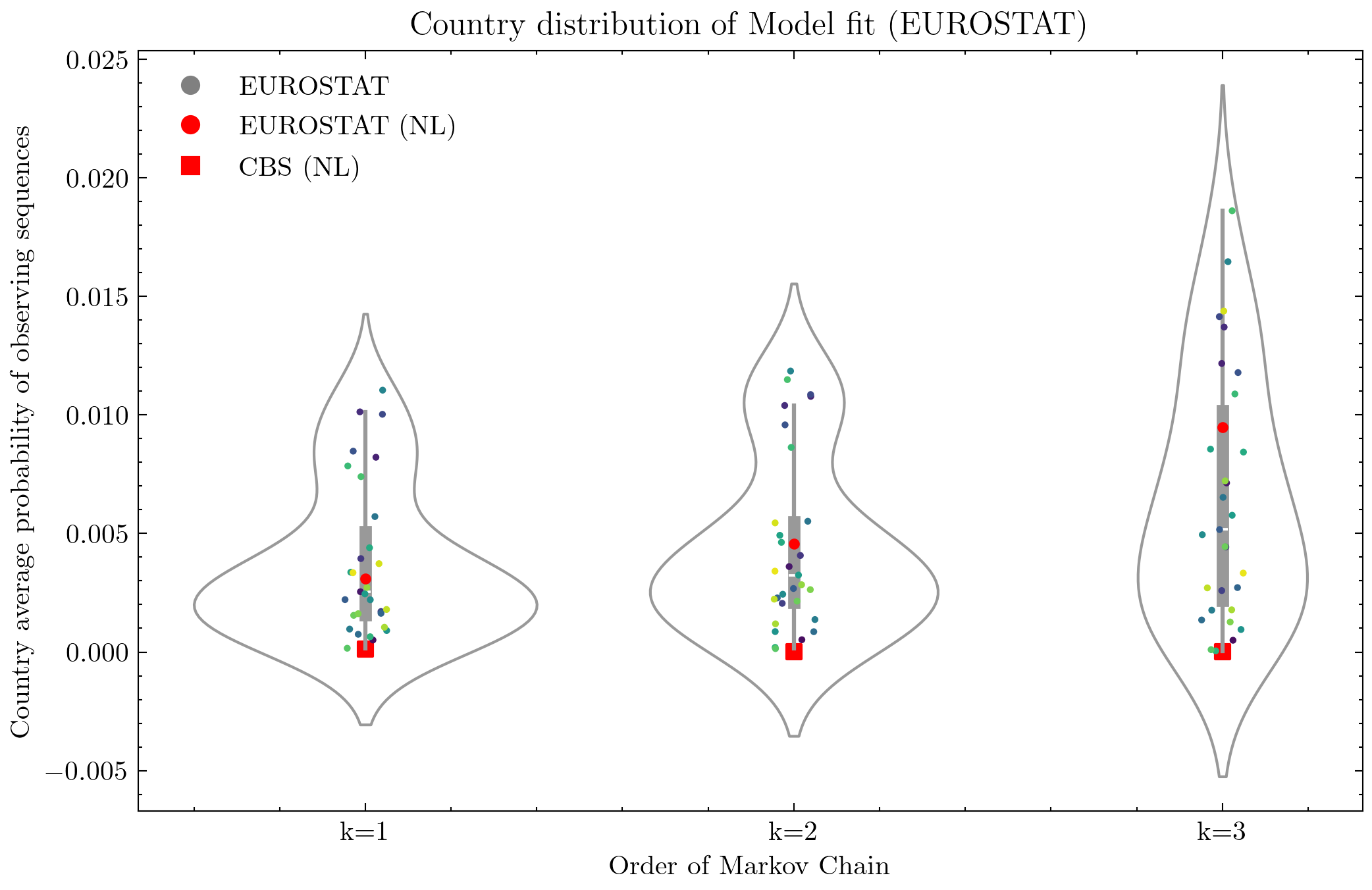}
    \caption{Given a Markov estimation for various countries we measure the probability of observing a sample of 5000 randomly selected household sequences given the transition matrix. While a larger memory length does result in better model performance, the differences are not big. Results from two different Dutch datasets vary significantly, indicating that between-country differences of this estimator are of a similar scale as two datasets from the same countries. }
    \label{fig:memory}
\end{figure}

Figure~\ref{fig:memory} compares the country-averaged probability of observing sequences for Markov chains of order $k=1$, $k=2$, and $k=3$ across European countries (for more detail consult table~\ref{tab:memory_length}). The plot includes results from EUROSTAT data (denoted by circles) and Dutch data from two distinct sources: EUROSTAT (highlighted by red circles) and CBS (represented by red squares).

From the figure, we observe that while increasing the order of the Markov chain from $k=1$ to $k=2$ or $k=3$ slightly improves the probability of observing sequences, the improvement is marginal. This suggests that while higher-order dependencies exist, they contribute minimally to the overall model fit, indicating that a first-order Markov chain may suffice for practical purposes. Notably, the data cross-validation using CBS data shows substantial variation, implying that discrepancies between data sources can be significant and must be carefully considered.

The computation of sequence probabilities involves assessing how well the Markov chain model of a given order predicts observed sequences in the dataset. Specifically, for each order $k$, the probability of observing a sequence of states is determined based on the estimated transition matrix $P_k$. The performance of the model is evaluated by comparing the likelihoods of the sequences generated under these matrices. This approach ensures that each memory length is assessed on the basis of its predictive capability, supporting the conclusion that higher-order Markov chains provide limited additional benefit over simpler first-order models.
\begin{table}[h!]
\centering
    \begin{tabular}{lrrr}
    \toprule
    \textbf{Country} & \textbf{\(k = 1\)} & \textbf{\(k = 2\)} & \textbf{\(k = 3\)} \\
    \midrule
    FR & 0.000508 & 0.000519 & 0.000498 \\
    SE & 0.002544 & 0.003605 & 0.007132 \\
    DK & 0.008214 & 0.010782 & 0.012171 \\
    EL & 0.010129 & 0.010398 & 0.013706 \\
    RS & 0.003941 & 0.004073 & 0.004430 \\
    PL & 0.001709 & 0.002053 & 0.002592 \\
    AT & 0.010027 & 0.010860 & 0.014143 \\
    NL & 0.003090 & 0.004556 & 0.009488 \\
    CH & 0.008472 & 0.009581 & 0.011786 \\
    MT & 0.002210 & 0.002680 & 0.005166 \\
    NO & 0.001634 & 0.002283 & 0.002719 \\
    LT & 0.000748 & 0.000858 & 0.001353 \\
    CZ & 0.005714 & 0.005516 & 0.006524 \\
    LU & 0.000968 & 0.001371 & 0.001766 \\
    CY & 0.011044 & 0.011851 & 0.016463 \\
    LV & 0.002204 & 0.002437 & 0.004949 \\
    BG & 0.000908 & 0.000862 & 0.000951 \\
    BE & 0.002453 & 0.003247 & 0.005769 \\
    UK & 0.003365 & 0.004924 & 0.008556 \\
    IT & 0.004398 & 0.004628 & 0.008434 \\
    PT & 0.000646 & 0.000197 & 0.000050 \\
    HR & 0.007848 & 0.008630 & 0.010884 \\
    IE & 0.007396 & 0.011489 & 0.018610 \\
    SK & 0.000160 & 0.000145 & 0.000101 \\
    RO & 0.001548 & 0.002147 & 0.004445 \\
    IS & 0.001617 & 0.002632 & 0.001268 \\
    HU & 0.002724 & 0.002839 & 0.007219 \\
    EE & 0.001046 & 0.001190 & 0.001781 \\
    FI & 0.001795 & 0.002231 & 0.002706 \\
    SI & 0.003727 & 0.005448 & 0.014381 \\
    ES & 0.003347 & 0.003412 & 0.003330 \\
    \bottomrule
    \end{tabular}
    \caption{Probability values for memory lengths \( k = 1 \), \( k = 2 \), and \( k = 3 \) across countries}
    \label{tab:memory_length}
\end{table}

\section{Error Estimation}
\label{errorest}
\begin{figure}[H]
\centering
\includegraphics[width=0.8\linewidth]{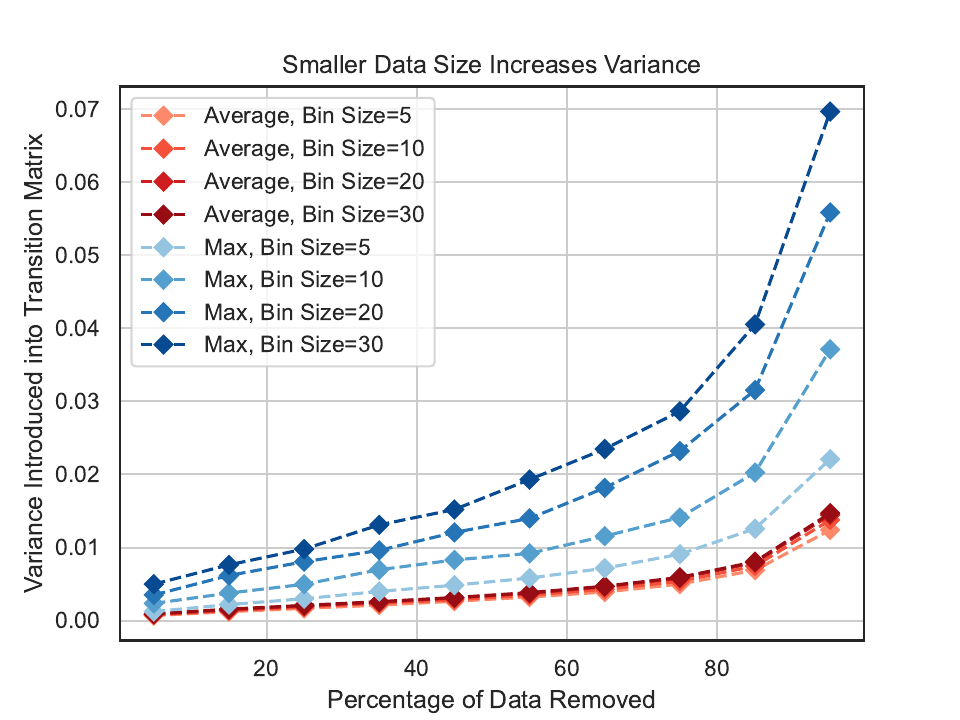}
\caption{\textbf{Overall error sensitivity to data.} Effect to average and maximum error of omitting increasing amount of data from the matrix estimation.}
\label{fig:overall_error}
\end{figure}

\begin{figure}[H]
\centering
\includegraphics[width=0.8\linewidth]{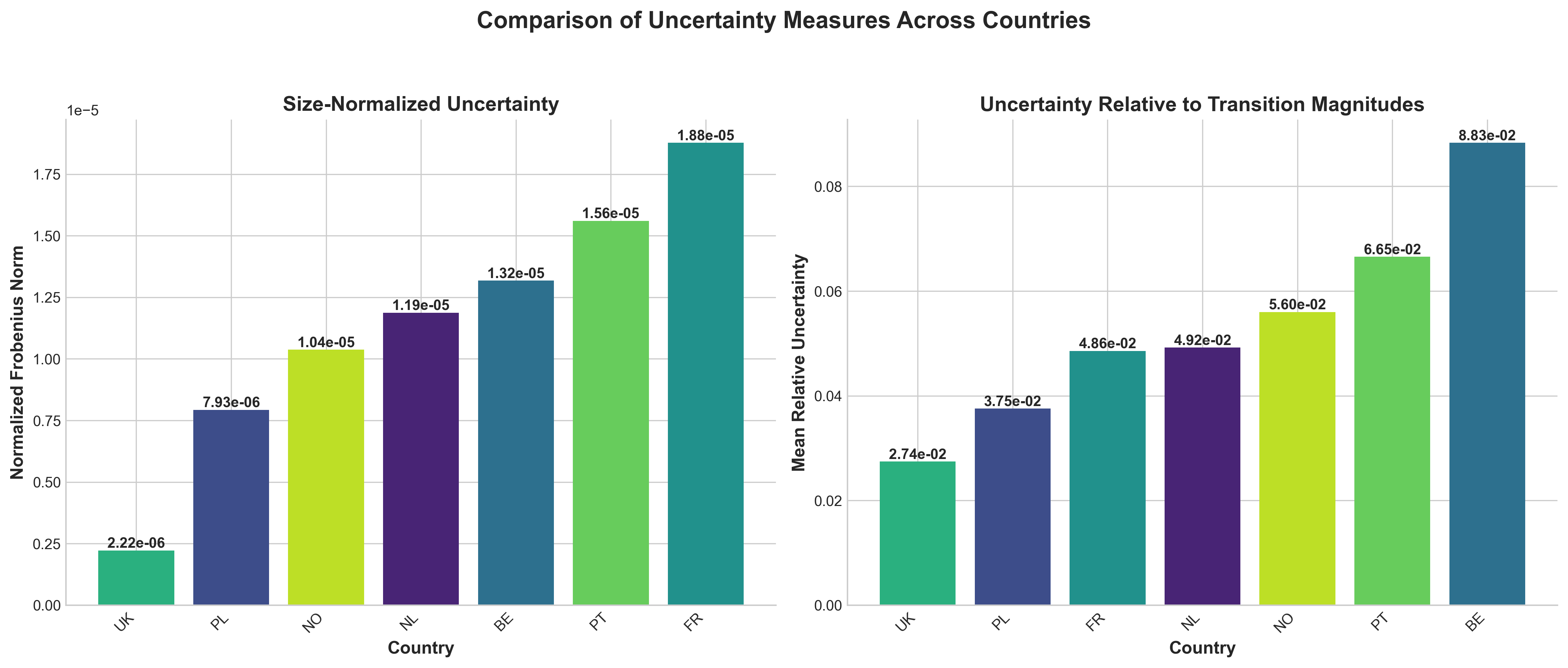}
\caption{\textbf{Country specific error estimates.} Error estimation using bootstrapping omitting 10\% of data. (left) the Frobenius norm of the matrix of standard deviations. (right) mean relative standard deviation of entries relative to entry size.}
\label{fig:errorest}
\end{figure}

\section{Key Metrics across eurozone}

\begin{table}[h!]
\centering
\caption{Traditional Poverty Measures vs.~Economic Landscapes}
\begin{tabular}{|p{3.7cm}|p{3cm}|p{5cm}|}
\hline
\textbf{Feature} & \textbf{Traditional} & \textbf{Economic Landscape} \\
\hline
What it shows & Who is poor? & Where traps form? How to escape? \\
\hline
Time dimension & Static snapshot & Dynamic trajectories over years \\
\hline
Multidimensionality & Counted separately & Synergistic interactions captured \\
\hline
Trap detection & No & Yes (visualized as valleys) \\
\hline
Escape time & Unknown & Quantified precisely \\
\hline
Recovery after shock & Uses GDP proxy & Directly measured \\
\hline
Policy guidance & Vague & Specific pathways identified \\
\hline
Cross-country comparison & Difficult & Direct landscape comparison \\
\hline
Resilience measurement & Not available & Mixing time \& recovery time metrics \\
\hline
\end{tabular}
\label{tab:comparemeasures}
\end{table}

\begin{table}[h!]
\centering
\caption{\textbf{Poverty and mobility metrics mapped to landscape framework properties.} Each row represents a distinct poverty or mobility metric; columns represent key properties emerging from the research questions in the Introduction.}
\label{tab:metrics_properties}
\small
\begin{tabularx}{\linewidth}{|p{1.5cm}|p{1.5cm}|p{2cm}|p{2cm}|p{1.9cm}|p{2cm}|}
\hline
\textbf{Metric} & \textbf{Dynamics} & \textbf{Dimension} & \textbf{Heterogeneity \& Traps} & \textbf{Resilience} & \textbf{ Policy} \\
\hline
\textbf{Income Poverty}\cite{foster2010foster} & Captures income-period only; cannot distinguish chronic vs transient & Uni-dimensional; ignores health-education interactions & No-treats all below-threshold as uniform & No-static measure cannot quantify recovery & Limited---identifies only income threshold crossing \\
\hline
\textbf{MPI}\cite{alkire2013multidimensional} & Aggregate snapshot; does not measure duration in deprivation & Multi-dimensional; captures joint deprivation but not threshold effects & Partial-identifies who is multiply deprived but not trapped trajectories & No-measures current state, not recovery dynamics & Moderate-shows intervention priorities by dimension \\
\hline
\textbf{AROPE}\cite{eurostat_arop_eu_silc} & Temporal component (12-month reference); distinguishes current vs prior year & Multi-dimensional (income, employment, education); limited synergy capture & Partial-relative poverty allows country comparisons but not trap identification & Limited-annual snapshot cannot quantify post-shock recovery & Moderate-identifies populations at risk, not drivers of persistence \\
\hline
\textbf{PGR}\cite{foster2010foster} & Temporal only if applied repeatedly; single-period measure & Uni-dimensional (income depth); no interaction effects & Shows severity but not persistence; cannot distinguish chronic poor from temporarily poor & Static; cannot measure shock adaptation & Low-shows only gap magnitude, not escape pathways \\
\hline
\textbf{SMI}\cite{shorrocks1976income} & Captures net upward/downward movement over periods & Single-dimension or aggregated; may miss dimension-specific stickiness & Reveals aggregate mobility patterns but not trap structure & Limited-does not separate shock response from structural dynamics & Moderate-shows aggregate movement but not optimal pathways \\
\hline
\textbf{IGE}\cite{becker1986human} & Long-term (life-course); captures persistence across generations & Typically uni-dimensional (income); recent extensions add education & Measures correlation of parent-child outcomes; high IGE = trap & Cannot directly measure shock response; implicit in long-term correlations & Moderate-identifies structural barriers but not intervention sequences \\
\hline
\end{tabularx}
\end{table}

\begin{table}[h!]
\centering
    \begin{tabular}{|c|c|c|}
    \hline
    \textbf{Country} & \textbf{1D Mixing Time} & \textbf{2D Mixing Time} \\
    \hline
    \ AT & 12.13 & 40.58 \\
    \ BE & 23.81 & 13.91 \\
    \ BG & 10.98 & 18.87 \\
    \ CH & 26.66 & 31.28 \\
    \ CY & 11.45 & 61.38 \\
    \ CZ & 11.10 & 459.62 \\
    \ DE & 20.90 & 176.83 \\
    \ DK & 45.90 & inf \\
    \ EE & 45.00 & 18.42 \\
    \ EL & 254.08 & 51.95 \\
    \ FI & 9.84 & 16.03 \\
    \ FR & 7.69 & 25.86 \\
    \ HR & 9.89 & 37.54 \\
    \ HU & 52.45 & 59.22 \\
    \ IE & 30.94 & 12.45 \\
    \ IS & 26.16 & 14.29 \\
    \ IT & 43.00 & 13.96 \\
    \ LT & 256.51 & 17.19 \\
    \ LU & 45.75 & 19.70 \\
    \ NL & 42.44 & 15.63 \\
    \ NO & 31.23 & 14.86 \\
    \ PL & 252.39 & 14.72 \\
    \ PT & 61.27 & 14.64 \\
    \ RO & 57.25 & 13.77 \\
    \ RS & 53.16 & 22.49 \\
    \ SE & 7.62 & 69.34 \\
    \ SI & 17.54 & 13.82 \\
    \ SK & 35.16 & 22.30 \\
    \ UK & 13.52 & 13.97 \\
    \hline
    \end{tabular}
    \caption{Combined 1D and 2D Mixing Times for Various Countries}
    \label{tab:combined_mixing_times}
\end{table}

\begin{table}[h!]
\centering
\begin{tabular}{l|ccc|ccc}
\hline
& \multicolumn{3}{c|}{Pre-COVID} & \multicolumn{3}{c}{During COVID} \\
Country & Mean & Lower & Upper & Mean & Lower & Upper \\
\hline
CH & 2.99 & 1.00 & 8.00 & 3.33 & 1.00 & 9.00 \\
BG & 2.98 & 1.00 & 8.00 & 2.88 & 1.00 & 8.00 \\
LU & 3.86 & 1.00 & 10.00 & 5.13 & 1.00 & 14.00 \\
HR & 4.16 & 1.00 & 12.00 & 3.97 & 1.00 & 11.00 \\
NL & 3.16 & 1.00 & 8.00 & 3.63 & 1.00 & 10.00 \\
BE & 3.41 & 1.00 & 9.00 & 3.36 & 1.00 & 9.00 \\
IT & 2.36 & 1.00 & 6.00 & 4.63 & 1.00 & 13.00 \\
FI & 2.09 & 1.00 & 5.00 & 2.47 & 1.00 & 6.00 \\
NO & 2.51 & 1.00 & 6.00 & 3.72 & 1.00 & 10.00 \\
LT & 2.28 & 1.00 & 6.00 & 2.54 & 1.00 & 6.00 \\
PL & 3.50 & 1.00 & 9.00 & 3.62 & 1.00 & 10.00 \\
LV & 1.82 & 1.00 & 4.00 & 2.33 & 1.00 & 6.00 \\
FR & 2.83 & 1.00 & 7.00 & 2.94 & 1.00 & 8.00 \\
CZ & 3.18 & 1.00 & 8.00 & 3.51 & 1.00 & 9.00 \\
SI & 2.65 & 1.00 & 7.00 & 3.63 & 1.00 & 10.00 \\
IE & 4.43 & 1.00 & 12.00 & 3.62 & 1.00 & 10.00 \\
ES & 2.00 & 1.00 & 5.00 & 2.50 & 1.00 & 6.00 \\
AT & 3.10 & 1.00 & 8.00 & 3.31 & 1.00 & 9.00 \\
SK & 3.04 & 1.00 & 8.00 & 4.75 & 1.00 & 13.00 \\
EL & 5.86 & 1.00 & 17.00 & 5.59 & 1.00 & 16.00 \\
RS & 2.44 & 1.00 & 6.00 & 2.54 & 1.00 & 6.00 \\
DK & 2.25 & 1.00 & 6.00 & 2.13 & 1.00 & 5.00 \\
HU & 3.33 & 1.00 & 9.00 & 3.81 & 1.00 & 10.00 \\
RO & 5.97 & 1.00 & 17.00 & 4.99 & 1.00 & 14.00 \\
PT & 1.13 & 1.00 & 2.00 & 1.79 & 1.00 & 4.00 \\
EE & 2.06 & 1.00 & 5.00 & 2.22 & 1.00 & 6.00 \\
MT & 2.12 & 1.00 & 5.00 & 1.85 & 1.00 & 4.00 \\
CY & 3.18 & 1.00 & 8.00 & 3.27 & 1.00 & 9.00 \\
\hline
\end{tabular}
\caption{Computed escape times in years from lowest income and health bin for estimated transition matrices before and during COVID and 90\% confidence interval.}
\label{tab:escape_times_comparison1}
\end{table}

\begin{table}[h!]
\centering
\begin{tabular}{l|ccc|ccc}
\hline
& \multicolumn{3}{c|}{Pre-COVID} & \multicolumn{3}{c}{During COVID} \\
Country & Mean & Lower & Upper & Mean & Lower & Upper \\
\hline
CH & 5.61 & 1.00 & 17.00 & 6.16 & 1.00 & 19.00 \\
BG & 3.66 & 1.00 & 11.00 & 3.54 & 1.00 & 10.00 \\
LU$^{*}$ & 6.22 & 1.00 & 20.00 & 9.41 & 1.00 & 30.00 \\
HR & 4.63 & 1.00 & 14.00 & 4.33 & 1.00 & 13.00 \\
NL & 7.72 & 1.00 & 24.00 & 8.79 & 1.00 & 27.00 \\
BE$^{*}$ & 5.95 & 1.00 & 19.00 & 7.01 & 1.00 & 21.00 \\
IT & 5.18 & 1.00 & 15.00 & 8.87 & 1.00 & 27.00 \\
FI & 3.31 & 1.00 & 9.00 & 4.75 & 1.00 & 14.00 \\
NO & 4.32 & 1.00 & 13.00 & 5.82 & 1.00 & 17.00 \\
LT & 3.25 & 1.00 & 9.00 & 3.28 & 1.00 & 9.00 \\
PL & 4.48 & 1.00 & 14.00 & 4.77 & 1.00 & 14.00 \\
LV & 3.00 & 1.00 & 8.00 & 3.30 & 1.00 & 9.00 \\
FR & 4.34 & 1.00 & 13.00 & 4.96 & 1.00 & 15.00 \\
CZ & 3.96 & 1.00 & 12.00 & 4.56 & 1.00 & 13.00 \\
SI & 4.36 & 1.00 & 12.00 & 5.50 & 1.00 & 16.00 \\
IE & 8.13 & 1.00 & 25.00 & 6.59 & 1.00 & 20.00 \\
ES & 4.04 & 1.00 & 12.00 & 4.93 & 1.00 & 15.00 \\
AT & 4.95 & 1.00 & 15.00 & 5.72 & 1.00 & 17.00 \\
SK & 3.78 & 1.00 & 12.00 & 6.88 & 1.00 & 21.00 \\
EL$^{*}$ & 7.92 & 1.00 & 26.00 & 7.40 & 1.00 & 24.00 \\
RS & 3.38 & 1.00 & 10.00 & 3.67 & 1.00 & 11.00 \\
DK & 4.17 & 1.00 & 13.00 & 3.92 & 1.00 & 12.00 \\
HU & 4.27 & 1.00 & 13.00 & 4.65 & 1.00 & 14.00 \\
RO$^{*}$ & 6.49 & 1.00 & 22.00 & 4.65 & 1.00 & 17.00 \\
PT & 1.57 & 1.00 & 3.00 & 2.06 & 1.00 & 5.00 \\
EE & 3.20 & 1.00 & 9.00 & 3.29 & 1.00 & 9.00 \\
MT & 3.92 & 1.00 & 11.00 & 3.51 & 1.00 & 10.00 \\
CY & 5.04 & 1.00 & 15.00 & 4.17 & 1.00 & 13.00 \\
\hline
\multicolumn{7}{l}{$^{*}$ Countries with exceptionally wide confidence intervals relative to their means.} \\
\end{tabular}
\caption{Computed escape times in years from lowest two income and health bin for estimated transition matrices before and during COVID and 90\% confidence interval.}
\label{tab:escape_times_comparison2}
\end{table}

\begin{table}[h!]
    \centering
    \begin{tabular}{lccc}
        \toprule
        Country & Pre-COVID Curl & During-COVID Curl & Delta Curl \\
        \midrule
        AT & 0.000833 & 0.001641 & 0.000808 \\
        BE & 0.001236 & 0.001679 & 0.000443 \\
        BG & 0.001665 & 0.001767 & 0.000102 \\
        CH & 0.002897 & 0.002353 & -0.000544 \\
        CY & 0.069017 & 0.031863 & -0.037154 \\
        CZ & 0.001811 & 0.001904 & 0.000093 \\

        DK & 0.002758 & 0.002910 & 0.000153 \\
        EE & 0.054049 & 0.027883 & -0.026166 \\
        EL & 0.002838 & 0.001610 & -0.001227 \\
        ES & 0.000929 & 0.000773 & -0.000156 \\
        FI & 0.149545 & 0.135407 & -0.014138 \\
        FR & 0.001341 & 0.001280 & -0.000062 \\
        HR & 0.001009 & 0.001134 & 0.000125 \\
        HU & 0.000831 & 0.000876 & 0.000045 \\
        IE & 0.002953 & 0.003231 & 0.000277 \\
        IT & 0.045488 & 0.031149 & -0.014339 \\
        LT & 0.151680 & 0.055522 & -0.096158 \\
        LU & 0.000679 & 0.000000 & -0.000679 \\
        LV & 0.118068 & 0.016695 & -0.101373 \\
        MT & 0.067912 & 0.058241 & -0.009671 \\
        NL & 0.001311 & 0.001388 & 0.000076 \\
        NO & 0.048391 & 0.029691 & -0.018700 \\
        PL & 0.049648 & 0.044621 & -0.005026 \\
        PT & 0.122711 & 0.124209 & 0.001499 \\
        RO & 0.070250 & 0.055045 & -0.015205 \\
        RS & 0.001335 & 0.001608 & 0.000272 \\
       
        SI & 0.000839 & 0.002363 & 0.001524 \\
        SK & 0.002311 & 0.002228 & -0.000084 \\
     
        \bottomrule
    \end{tabular}
    \caption{Total curl comparison across countries before and during the COVID-19 period. Curl is computed as the sum of absolute net probability flows between all state pairs, defined as $\sum_{i < j} |\pi_i p_{ij} - \pi_j p_{ji}|$, where $\pi$ is the stationary distribution and $p_{ij}$ is the observed transition probability. Higher values indicate stronger violations of time-reversibility. Values are small compared to baseline noise (e.g., by binning) }
    \label{tab:curl}
\end{table}

\section{EU Landscapes}
\label{SI:landscape}
\begin{figure}[H]
\centering
\includegraphics[width=0.95\linewidth]{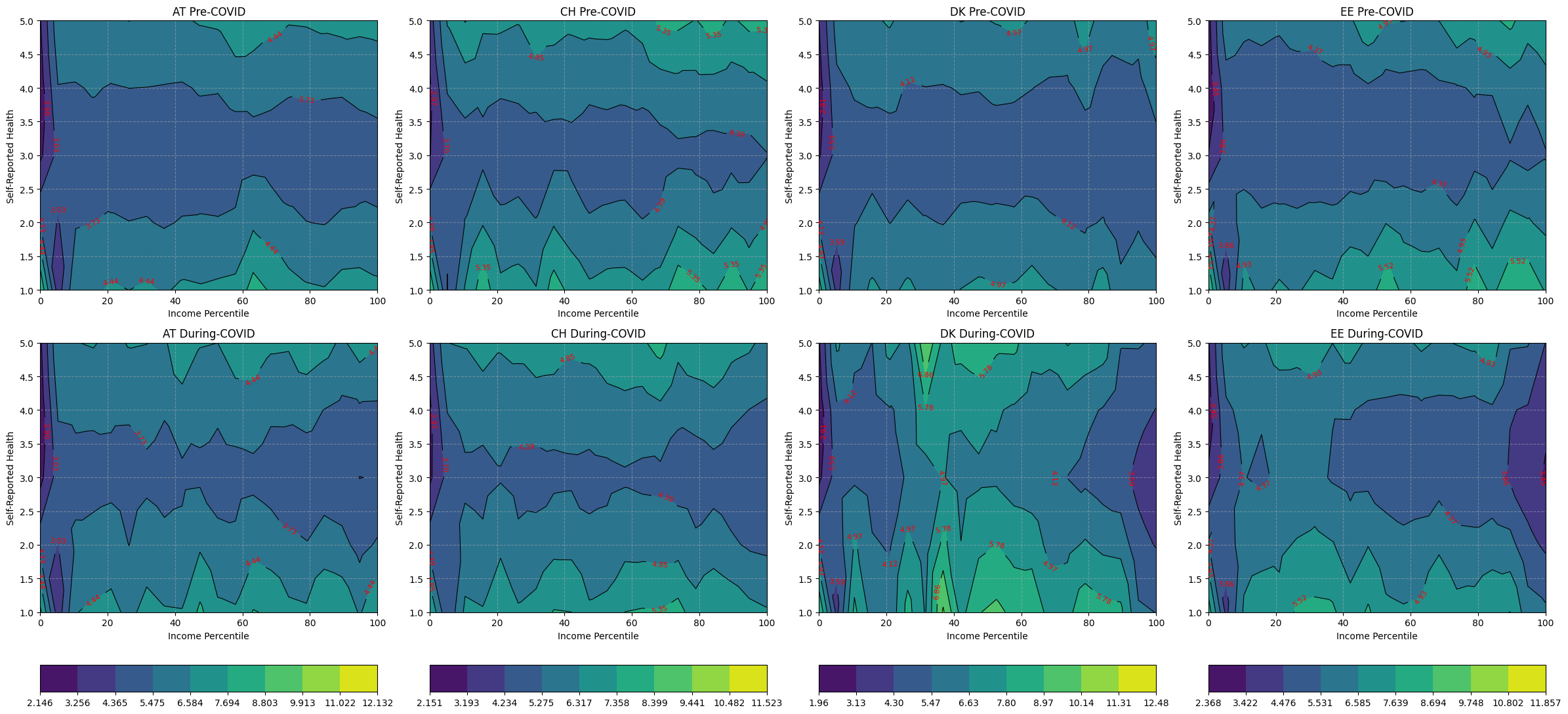}
\caption{ COVID-19 impacts on welfare landscapes across European countries (Austria, Switzerland, Denmark, Estonia). Potential energy maps for health and income dimensions before (top) and during (bottom) the pandemic for selected countries. Contour lines show mobility barriers, revealing increased obstacles for lower-income groups while higher-income households maintained their positions.}
\label{fig:landscape1}
\end{figure}

\begin{figure}[H]
\centering
\includegraphics[width=0.95\linewidth]{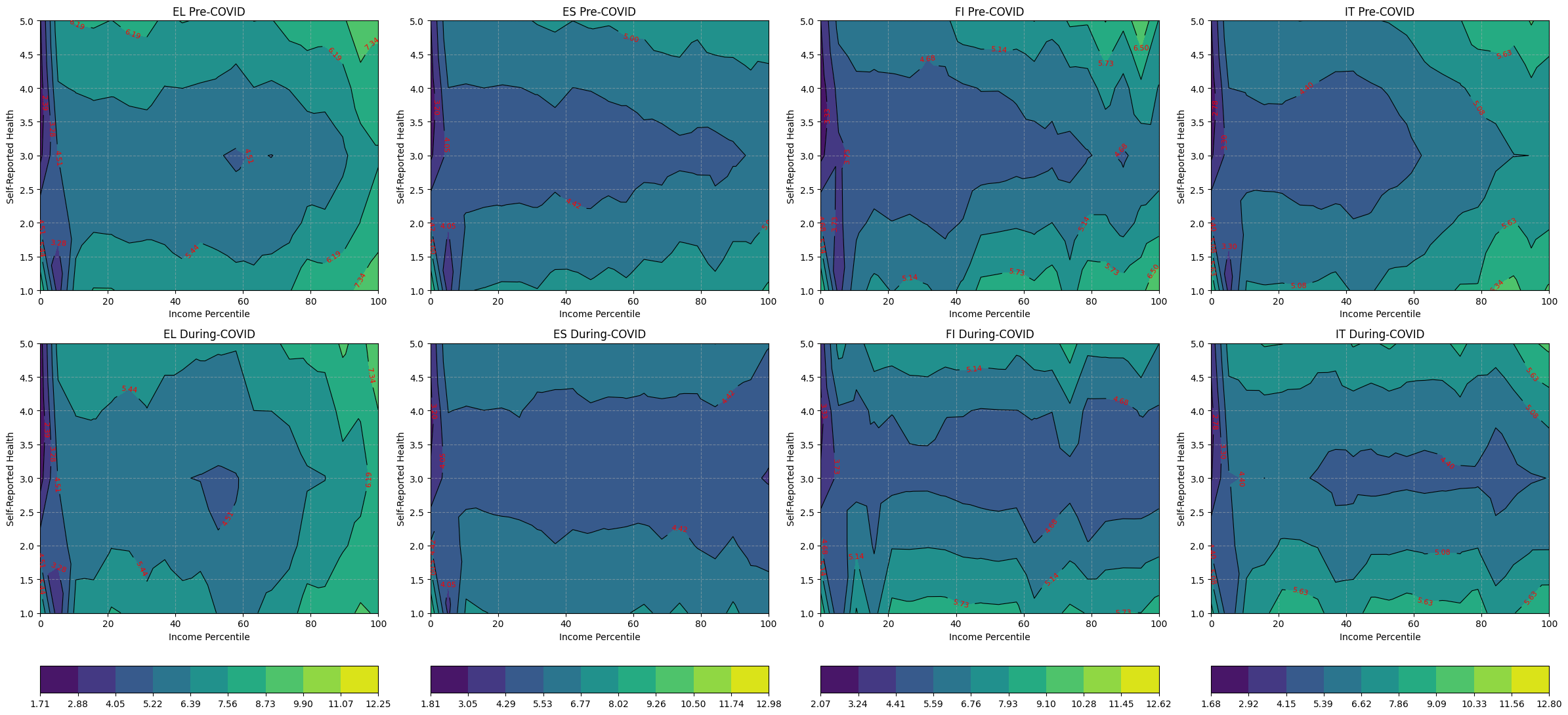}
\caption{COVID-19 impacts on welfare landscapes across European countries (Greece, Spain, Finland, Italy). Potential energy maps for health and income dimensions before (top) and during (bottom) the pandemic for selected countries. Contour lines show mobility barriers, revealing increased obstacles for lower-income groups while higher-income households maintained their positions.}
\label{fig:landscape2}
\end{figure}

\begin{figure}[H]
\centering
\includegraphics[width=0.95\linewidth]{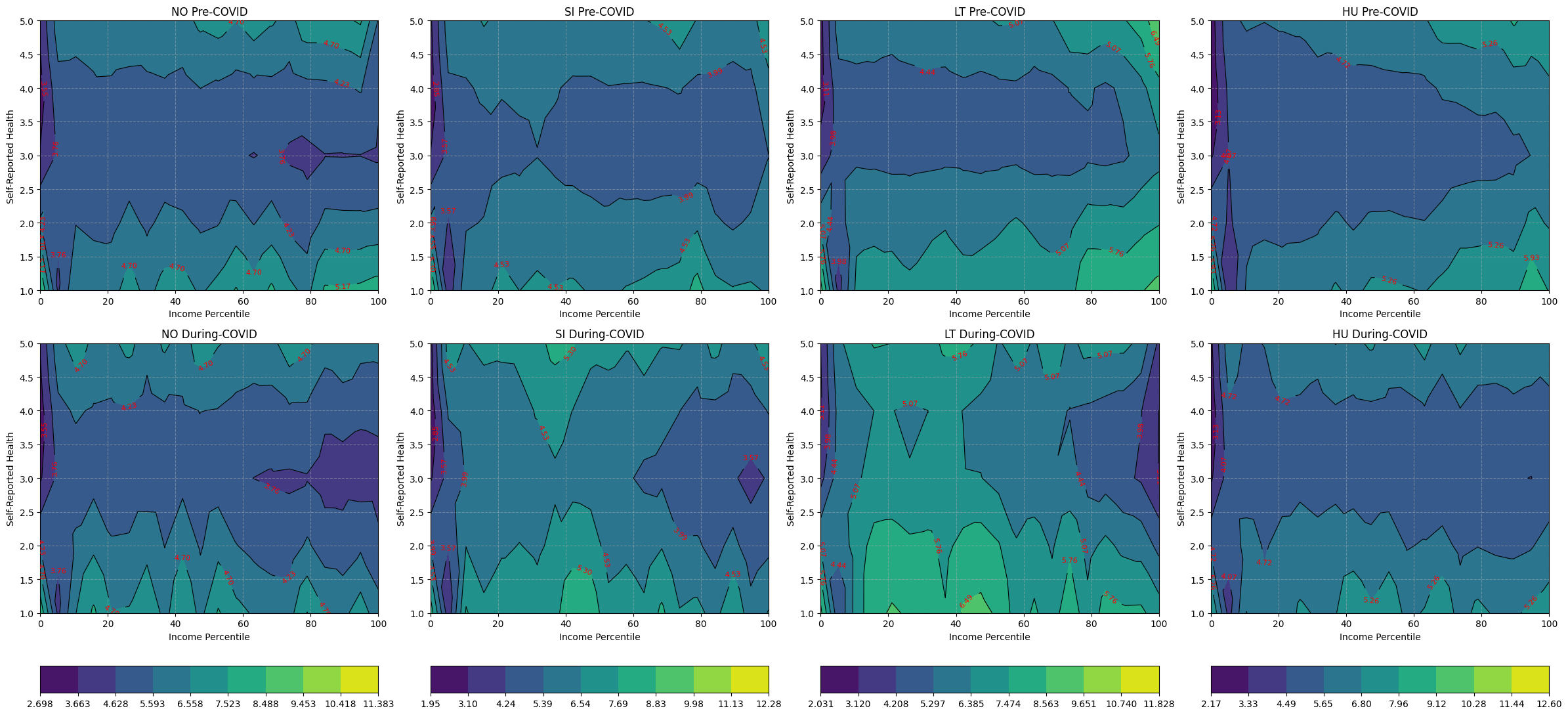}
\caption{ COVID-19 impacts on welfare landscapes across European countries (Norway, Slovenia, Lithuania, Hungary). Potential energy maps for health and income dimensions before (top) and during (bottom) the pandemic for selected countries. Contour lines show mobility barriers, revealing increased obstacles for lower-income groups while higher-income households maintained their positions.}
\label{fig:landscape3}
\end{figure}

\section{Data description}
\label{data}
\begin{table}[h!]
\centering
\begin{tabular}{l|lll|llll}
\toprule
& \multicolumn{3}{c|}{Years} & \multicolumn{4}{c}{Statistics} \\
 & 2006 & 2013 & 2021 & 25\% & 50\% & 75\% & max \\
\midrule
FR & 23.0 ± 18.0 & 24.0 ± 22.0 & 28.0 ± 23.0 & 10.0 & 20.0 & 30.0 & 1075.0 \\
SE & 23.0 ± 19.0 & 29.0 ± 25.0 & 32.0 ± 29.0 & 10.0 & 25.0 & 37.0 & 824.0 \\
DK & 34.0 ± 24.0 & 43.0 ± 32.0 & 51.0 ± 45.0 & 16.0 & 39.0 & 54.0 & 10938.0 \\
EL &  & 17.0 ± 14.0 & 16.0 ± 11.0 & 9.0 & 14.0 & 20.0 & 359.0 \\
RS &  &  & 7.0 ± 5.0 & 4.0 & 5.0 & 7.0 & 83.0 \\
PL & 5.0 ± 5.0 & 8.0 ± 7.0 & 11.0 ± 8.0 & 4.0 & 7.0 & 10.0 & 178.0 \\
AT & 24.0 ± 18.0 & 29.0 ± 32.0 & 35.0 ± 31.0 & 11.0 & 23.0 & 38.0 & 1197.0 \\
NL & 28.0 ± 26.0 & 33.0 ± 31.0 & 35.0 ± 34.0 & 11.0 & 27.0 & 43.0 & 1024.0 \\
CH &  &  & 64.0 ± 82.0 & 19.0 & 48.0 & 78.0 & 4266.0 \\
MT &  & 16.0 ± 15.0 & 23.0 ± 19.0 & 9.0 & 15.0 & 22.0 & 282.0 \\
NO & 31.0 ± 26.0 & 46.0 ± 45.0 &  & 12.0 & 36.0 & 57.0 & 1416.0 \\
LT & 4.0 ± 4.0 & 6.0 ± 5.0 & 13.0 ± 11.0 & 3.0 & 5.0 & 9.0 & 98.0 \\
CZ & 6.0 ± 4.0 & 10.0 ± 7.0 & 13.0 ± 10.0 & 5.0 & 8.0 & 12.0 & 231.0 \\
LU & 41.0 ± 33.0 & 47.0 ± 36.0 & 59.0 ± 49.0 & 22.0 & 36.0 & 60.0 & 1804.0 \\
CY & 17.0 ± 18.0 & 21.0 ± 23.0 & 22.0 ± 31.0 & 7.0 & 14.0 & 24.0 & 958.0 \\
LV &  & 6.0 ± 7.0 & 12.0 ± 12.0 & 3.0 & 5.0 & 9.0 & 159.0 \\
BG &  & 4.0 ± 3.0 & 6.0 ± 7.0 & 2.0 & 3.0 & 5.0 & 143.0 \\
BE & 29.0 ± 23.0 & 34.0 ± 22.0 & 35.0 ± 32.0 & 18.0 & 29.0 & 40.0 & 1880.0 \\
UK & 32.0 ± 30.0 & 30.0 ± 33.0 &  & 14.0 & 24.0 & 38.0 & 1659.0 \\
IT &  & 22.0 ± 18.0 & 24.0 ± 21.0 & 10.0 & 19.0 & 28.0 & 2473.0 \\
DE &  &  &  & 13.0 & 28.0 & 44.0 & 559.0 \\
PT &  & 13.0 ± 12.0 &  & 8.0 & 11.0 & 16.0 & 336.0 \\
HR &  & 9.0 ± 6.0 & 12.0 ± 8.0 & 6.0 & 8.0 & 13.0 & 185.0 \\
IE & 28.0 ± 25.0 & 34.0 ± 31.0 & 35.0 ± 34.0 & 12.0 & 25.0 & 43.0 & 934.0 \\
SK & 4.0 ± 10.0 & 8.0 ± 5.0 & 10.0 ± 6.0 & 5.0 & 7.0 & 9.0 & 212.0 \\
RO &  & 4.0 ± 2.0 & 10.0 ± 5.0 & 3.0 & 4.0 & 7.0 & 50.0 \\
IS & 33.0 ± 30.0 & 24.0 ± 22.0 &  & 11.0 & 24.0 & 40.0 & 1191.0 \\
HU & 5.0 ± 7.0 & 6.0 ± 5.0 & 7.0 ± 5.0 & 2.0 & 5.0 & 7.0 & 162.0 \\
EE & 5.0 ± 4.0 & 9.0 ± 7.0 & 14.0 ± 14.0 & 4.0 & 7.0 & 11.0 & 344.0 \\
FI & 21.0 ± 22.0 & 28.0 ± 26.0 & 33.0 ± 31.0 & 7.0 & 23.0 & 36.0 & 606.0 \\
SI & 11.0 ± 11.0 & 15.0 ± 13.0 & 18.0 ± 16.0 & 5.0 & 12.0 & 19.0 & 235.0 \\
ES & 16.0 ± 12.0 & 18.0 ± 17.0 & 20.0 ± 19.0 & 7.0 & 14.0 & 23.0 & 391.0 \\
\bottomrule
\end{tabular}
\caption{ Mean income and standard deviation by country for selected years (in thousands, PPP-adjusted) of EU-SILC dataset.}
\end{table}

We use data from the European Union Statistics on Income and Living Conditions (EU-SILC), a harmonized microdata source coordinated by Eurostat and implemented by national statistical offices. Income is collected at both the household and individual levels and includes a wide range of components such as employment income, self-employment earnings, property income, pensions, and social transfers. These are reported both gross and net, typically referring to the previous calendar year, and are gathered via interviews and administrative registers depending on the country. Here, we used total net income. Self-reported health is measured using the Minimum European Health Module (MEHM), which consists of three standardized questions: self-perceived general health (rated on a five-point scale), the presence of longstanding health problems, and limitations in daily activities due to health issues. Here, we used the answers to the first question. These variables are subjective and consistently collected across countries. Educational attainment is recorded for individuals aged 16 and over, based on the highest level of education completed, and is classified according to the International Standard Classification of Education (ISCED) to ensure comparability across national systems.

\begin{table}[h]
    \centering
    \begin{tabular}{ll}
        \toprule
        \textbf{Country} & \textbf{Alpha-2 Code} \\
        \midrule
        Austria        & AT  \\
        Belgium        & BE  \\
        Bulgaria       & BG  \\
        Croatia        & HR  \\
        Cyprus         & CY  \\
        Czechia        & CZ  \\
        Denmark        & DK  \\
        Estonia        & EE  \\
        Finland        & FI  \\
        France         & FR  \\
        Germany        & DE  \\
        Greece         & GR  \\
        Hungary        & HU  \\
        Ireland        & IE  \\
        Italy          & IT  \\
        Latvia         & LV  \\
        Lithuania      & LT  \\
        Luxembourg     & LU  \\
        Malta          & MT  \\
        Netherlands    & NL  \\
        Poland         & PL  \\
        Portugal       & PT  \\
        Romania        & RO  \\
        Slovakia       & SK  \\
        Slovenia       & SI  \\
        Spain          & ES  \\
        Sweden         & SE  \\
        \bottomrule
    \end{tabular}
    \caption{ISO 3166-1 Alpha-2 Country Codes of EU Member States}
    \label{tab:eu_country_codes}
\end{table}

\end{appendices}

\end{document}